\documentclass[aps,prd,twocolumn,superscriptaddress,nofootinbib]{revtex4-2}

\usepackage{amsmath, amssymb, amsfonts}
\usepackage{graphicx}
\usepackage{bm}
\usepackage{hyperref}
\usepackage[utf8]{inputenc}
\usepackage{multirow}
\usepackage{graphicx}
\usepackage[export]{adjustbox}
\usepackage{physics}
\usepackage{hyperref}
\hypersetup{
    colorlinks=true,
    linkcolor=blue,
    filecolor=magenta,      
    urlcolor=cyan,
    }
\begin{document}

\title{(De)Coherence of a quantum system in an anti-de Sitter spacetime}

\author{Mandas Biswas}
\affiliation{School of Physical Sciences, Indian Association for the Cultivation of Science, 2A \& 2B Raja S.C. Mullick Road, Kolkata-700032, India.}

\author{Anupam Mazumdar}
\affiliation{Van Swinderen Institute for Particle Physics and Gravity, University of Groningen, 9747AG, Groningen, The Netherlands.}
\date{\today}

\begin{abstract}
In this paper, we study the coherence/decoherence of a quantum harmonic oscillator in anti-de Sitter (AdS) spacetime by quantising the graviton in a curved background with a nontrivial boundary condition. In the quantum-field-theory framework, we obtain a master equation by tracing away the gravitational field
at the leading order in ${\cal O}(G,~ 1/\omega^{2})$, where $\omega$ is related to the trapped frequency of the harmonic oscillator. We will proceed with a semi-Markovian analysis to compute the Lindbladian equation and estimate the decoherence rate and selection rules at the leading order, which come in two kinds: one responsible for the local interaction between matter and graviton, and the other related to the AdS global curvature. The latter determines the recoherence of the quantum system for a certain choice of the trapped harmonic oscillator's frequency and the global curvature. We also demonstrate that we recover the flat spacetime limit by taking appropriate approximations.

\end{abstract}

\maketitle

\section{Introduction}

A quantum system in a gravitational background can serve as the simplest toy model for quantum gravity if gravity is treated on a par with quantum mechanics in that background. For instance, a matter-wave interferometer~\cite{Arndt:2022} interacting with a graviton in a Minkowski background~\cite{Gupta:1954zz} would be an ideal system for demonstrating a nontrivial quantum correlation, known as quantum entanglement~\cite{Horodecki:2009zz}, see
Ref.~\cite{Rufo:2024ulr,Mazumdar:2026ewn}. 
The matter-graviton vertex gives rise to this tripartite entanglement between the two helicity states of the massless graviton and the matter around a Minkowski background. 
The entanglement is bona fide quantum correlation which has no classical counterpart~\cite{Horodecki:2009zz}. 

A quantum matter-wave interferometer emits gravitational waves; a simple example is a trapped harmonic oscillator~\cite{Toros:2020krn}, similar to its classical analog~\cite{Gasperini:1986}. Similarly, quantum matter emits gravitational waves that follow distinct selection rules~\cite{Toros:2020krn}, and also gives rise to gravitationally induced decoherence, see~\cite{Anastopoulos:2013zya,Blencowe:2012mp,DeLorenci:2014vwa,Oniga:2015lro,Oniga:2017pyq,Parikh:2020nrd,Parikh:2020fhy,Danielson:2022tdw,Danielson:2022sga,Wilson-Gerow:2024ljx,biggs2024comparingdecoherenceeffectsblack}.

Recently, an experimental proposal has also been devised to witness the quantum nature of gravity in a lab~\cite{Bose:2017nin,Marshman:2019sne,Bose:2022uxe}, see also~\cite{marletto2017gravitationally}. In particular, witnessing a virtual excitation of a graviton in an optomechanical setup, which is a quantum counterpart of the light-bending experiment due to gravity, i.e., graviton-induced entanglement~\cite{Biswas:2022qto,Biswas:2026pem}.

AdS spacetime has been the center of the study for understanding quantum theories of gravity since the discovery of AdS/CFT correspondence \cite{Maldacena_1999}, and the consequent realization of the holographic principle \cite{witten1998antisitterspaceholography}.
Following the Ryu-Takyanagi conjecture \cite{Ryu_2006, Ryu_2006_1}, various attempts have been made for formulating holographic quantum gravity in terms of quantum information to understand how gravitational field (spacetime) may be an emergent phenomena - some important papers in these directions are \cite{penington2020entanglementwedgereconstructioninformation, Almheiri_2015, PhysRevD.111.066021}; for a comprehensive review see \cite{takayanagi2025emergentholographicspacetimequantum}. 

AdS gravitons have been studied before, mainly in the context of propagators and the question of whether they have smooth limits to flat-space graviton propagators \cite{D_Hoker_1999, lzqw-sp76}. Decoherence and recoherence calculations using open quantum system methods have been applied for gravitational systems mainly in the context of the black hole information problem \cite{Anglin_1995, biggs2024comparingdecoherenceeffectsblack}.

Here, our aim is very different; we now wish to understand the quantum features of a graviton in an anti-de Sitter (AdS) background. We want to understand how a quantum oscillator emits gravitational waves, and how it loses its coherence. However, AdS has a nontrivial boundary, which makes the analysis extremely interesting. The non-trivial reflective boundary also yields a recoherence rate that can now be estimated. We will study the decoherence of a quantum oscillator by following the Lindblad master equation in an AdS background. Strictly speaking, the setup is not Markovian, but under the assumption that the oscillator's trapping frequency is large, this means the time and length scales are smaller than any other length scales in the setup, which helps us to effectively capture the leading-order physics even at the zeroth level, due to nontrivial boundary conditions and a non-vanishing curved background. Therefore, our goal is very modest; we wish to understand the decoherence of a quantum system in a finite gravitational system, viz., anti-de Sitter spacetime. We will work in the global $(1+3)$ AdS coordinates in its static chart $(t,r,\theta,\phi)$.

We will first discuss the properties of the global AdS background, and then those of a massive free particle in an AdS background. The geometry provides a trapping frequency; this part of the computation is strictly classical. We then quantise gravity in the linearised limit in the presence of the quantum harmonic oscillator in global AdS. We will then study the coherence and decoherence of the quantum harmonic oscillator using the semi-Markovian technique and identify the selection rules which will protect the coherence of the quantum system. 

\section{\texorpdfstring{Global anti-de Sitter space}{The Classical Set-Up of the Formalism in GAdS 4}}

The vacuum Einstein-Hilbert action with a $-$ve cosmological constant is given by~\footnote{Unless otherwise specified, we use the $(-+++)$ metric signature. Also, similarly, unless otherwise mentioned, Greek indices run from 0 to 3, and Latin indices run from 1 to 3.}
\begin{eqnarray}   
    S = -\frac{1}{16\pi}\int d^4x\sqrt{-g}(R - 2\Lambda) \nonumber \\    -\frac{1}{8\pi}\int_{\partial(\text{boundary})}d^3x\sqrt{-\gamma}\Theta
\end{eqnarray}
where $\gamma$ is the metric on the boundary and $\Theta$ is the trace of the boundary's extrinsic curvature, as defined in Appendix \eqref{extrinsic}. We will not consider the latter term, as it is included only to make the variational problem well-defined at the boundary \cite{GibbonsHawking1977, BrownYork1993}, and our analysis in the paper will be conducted close to the centre of the AdS spacetime. 
The $(1+3)$-AdS in its static chart $(t, r, \theta,\phi)$ can be derived as a negative cosmological constant vacuum solution to the Einstein field equation, 
\begin{equation}
G^\mu_\nu + \Lambda\delta^\mu_\nu = 0,
\end{equation}
and is given by,
\begin{equation}
    ds^2 = -(1+\frac{r^2}{L^2})c^2dt^2 + (1+\frac{r^2}{L^2})^{-1}dr^2 + r^2d\Omega_2^2
    \label{GM}
\end{equation}
where, 
\begin{equation}
\Lambda = -\frac{3}{L^2}\,.
\end{equation}
This is the AdS curvature; this means that if we embed this metric in $\mathbb{R}^{2,3}$, then this metric will describe the hyperregion within the space given by the equation, 
\begin{equation}
-x_{-1}^2 - x_0^2 + x_1^2 + x_2^2 + x_3^2 = - L^2
\end{equation}
making this space globally hyperbolic. 
For no value of $r$, will the $g_{tt}$ or $g_{rr}$ components be zero in this chart, thus apparently having no coordinate singularity or a boundary like de-Sitter spacetime (the positive constant curvature solution). However, if we make a substitution 
\begin{equation}
r=L\sinh\rho\implies dr = L\cosh\rho d\rho
\end{equation}
and substitute this back in Eq.~\eqref{GM}, we get the $GAdS_4$-metric, given by,
\begin{equation}
    ds^2 = L^2 \left( -\cosh^2 \rho \, d\tau^2 + d\rho^2 + \sinh^2 \rho \, d\Omega_{2}^2 \right), t\in \mathbb{R}
    \label{AdSm}
\end{equation}
where, $\tau = c/L dt$ is the dimensionless global AdS time. We note here two things:
\begin{itemize}
\item We have "unwrapped" the time coordinate here. This will be more apparent if we look at the $GAdS_4$ parametrisation from the higher-dimensional embedding, viz., by substituting 
\begin{equation}
x_{-1} = L\cosh\rho\cos \tau,~~~~~~x_0 = L\cosh\rho\sin \tau
\end{equation}
with the 2-sphere coming from the $r=L\sinh\rho$, we arrive at the same metric with the identification that 
\begin{equation}
\tau \sim \tau + 2\pi.
\end{equation}
This results in \textit{closed timelike curves}, which is physically unacceptable. To get around this problem, we have mapped $\tau\longrightarrow t: S^1 \longrightarrow\mathbb{R}$. Also, the domain of $\rho$ runs from $[-\infty, +\infty]$, as $\cosh\rho > 0$ for all $\rho$. In general, for most part of the rest of the paper, we'll choose $\rho = 0$ as the centre of the $GAdS_4$ space. 

\item Secondly, as a result of the previous observation, we calculate the time required for a massless radial observer to reach the AdS horizon (or equivalently, the boundary). By taking $ds^2 = 0,~ d\Omega_2^2 = 0$ we obtain, 
\begin{equation}
    \cosh^2\rho ~d\tau^2 = d\rho^2\implies d\tau = \frac{d\rho}{\cosh\rho}\,.
    \label{timefornow}
\end{equation}
We integrate the RHS from $0$ to $\infty$, and we obtain,
\begin{equation}
    \tau = \int^\infty_0 \frac{d\rho}{\cosh\rho} = \frac{\pi}{2}\,.\nonumber
\end{equation}
Going back to the static observer time, we will have,
\begin{equation}
    t = \frac{\pi L}{2c}\,.
    \label{time}
\end{equation}
Thus, light will reach the AdS boundary in this "unwrapped" coordinate in a finite coordinate time.

\end{itemize}

However, if we look at the affine parametrisation, we can rewrite Eq.~\eqref{timefornow} as $\cosh^2\rho \dot{t}^2 = \dot{\rho}^2, \cosh^2\dot{t} = E$, where $E$ is the conserved energy associated with $t$-translation invariance of EQ.~\eqref{GM}, and $\dot~= d/d\upsilon$. We can use the second equation in the first one to get $\frac{d}{d\upsilon}\sinh\rho = \pm E$, where $\upsilon$ is the affine parameter. We integrate out in the affine parameter to obtain, for outgoing light rays (the plus sign), $\sinh\rho(\upsilon) = E(\upsilon - \upsilon_0)$. Thus, it will take infinite affine time to reach $\rho = \infty$. 

This is a strange feature of AdS, which we will explore in this paper - for an observer localised near the centre of the AdS spacetime, $\rho = 0$, an outgoing light-ray/gravitational wave reaches (and with appropriate boundary conditions, comes back) in finite coordinate time, but for the light-ray itself, it takes infinite time to reach the AdS horizon. The corresponding Penrose diagram can be found in section 39.3.2 in Ref.~\cite{BlauGR2025}. 

\subsection{Gravitational Radiation for Free Particles} 

In general, free point particles have no mass quadrupole moment by definition, and the lowest configuration with a nonzero quadrupole moment is a coupled harmonic oscillator. We will see how a harmonic oscillator behaves in $GAdS$ in the next section. However, we note that free particles always have periodic, closed trajectories (orbits) in AdS space. This section is entirely classical, and we do not quantise graviton.

To see that, we write down the action of a relativistic point particle moving non-relativistically in a general curved spacetime,
\begin{equation}
    S = -m\int d\tau = -m\int dt \sqrt{-g_{\mu\nu}\frac{dx^\mu}{dt}\frac{dx^\nu}{dt}}
    \label{PS}
\end{equation}
Rather than trying to substitute the metric components and extremising the action to get the Euler-Lagrange equations, we can actually solve for the dynamics of the particle in the embedding space with the particle constrained to move in the surface $x^Ax_A = L^2$, where $A = -1, 0, 1, 2, 3$, and $L$ is related to the cosmological constant in AdS. Thus introducing the Lagrange multiplier $\xi$ we have \eqref{PS} turn out to be,
\begin{equation}
    S[x^A, \xi] = -\int d\tau(\dot{x}_A\dot{x}^A + \xi(L^2 + x^Ax_A))
\end{equation}
The equations of motion become (once varying w.r.t. $\tau$ and another time, w.r.t. $\xi$) yields,
\begin{equation}
    \ddot{x}_A = -\xi x_A;~~~~~~~~ x_Ax^A + L^2 = 0
    \label{EOM}
\end{equation}
We can rescale the unphysical worldline coordinate $\tau$ by any positive quantity to identify with the time coordinate in the GAdS metric, therefore constraining us essentially to three values of $\xi = -1, 0, 1$. They correspond to spacelike, null and timelike particles, respectively. We will work with timelike particles, which correspond to $\xi = 1$. With this value of $\xi=1$ and replacing $\tau$ with $t$ in the first equation of \eqref{EOM}, we can see a set of solutions to be, 
\begin{eqnarray}
    x_1 &=& L\tan\rho_1\cos t\nonumber\\
    x_{-1} &=& L\sin t\nonumber\\
    x_{0} &=& \frac{L\cos t}{\cos\rho_1};~~~~~ x_i = 0\text{ }(\forall i\neq 1)
\end{eqnarray}
In this solution, we can clearly see that the particle is oscillating in the $\rho$-direction between $\pm\rho_1$, for $\rho_1 \ll 1$.

Since AdS is a maximally symmetric spacetime, it does not matter what we choose as the centre of the spacetime, as all points are equivalent up to some conformal transformation for all choices of $\rho$ by AdS isometry. Even all trajectories are equivalent under some symmetry transformations. We'll generally choose $\rho_1 \sim 0$ in the upcoming sections whenever we have to explicitly use any value of $\rho$, as the localised position of the oscillator. For other solutions, see \cite{KaplanAdSCFT}.  

All the solutions have oscillatory frequencies~\footnote{We have performed the calculation here for a free particle in $ GAdS$. The negativity of the cosmological constant induces a harmonic potential, and that frequency is given by this expression. For a harmonic oscillator with frequency $\omega_m$, this harmonic trap maps the frequency to an effective value. For more details on this, see section IIIA.} 
\begin{equation}
\omega = \frac{c}{L}
\end{equation} 
 The amplitude of the oscillations is given by: $A = L\tan\rho_1$, and $x(t) = L\tan\rho_1\cos(\omega t)$, and $y(t) = z(t) = 0$. We can calculate the quadrupole moment for a point mass using $$Q_{ij} = m(3x_ix_j - r^2\delta_{ij})$$ 
We note, $r^2 = x^2 = A^2\cos^2(\omega t) = \frac{A^2}{2}(1+\cos(2\omega t))$. Thus, we can have the classical Einstein quadrupole radiation power \cite{Maggiore2008},
\begin{align}
    \frac{dE}{dt} &= -\frac{G}{45c^2}\sum_{i, j}(\dddot{Q}_{ij})^2 = -\frac{32G}{15c^5}m^2A^4\omega^6\sin^2(2\omega t) \nonumber\\
    &= -\frac{32Gm^2c}{15L^2}\tan^4\rho_1\sin^2\left(\frac{2ct}{L}\right)
    \label{quad}
\end{align}
where $\omega=c/L$. We can easily check that in the long-time average limit over a period of oscillation, and noting $<\sin^2(2\omega t)> = {1}/{2}$, this reduces to the flat-space equation~\cite{Maggiore2008}.  

\section{\texorpdfstring{Linearised Quantum Gravity in $GAdS_4$}{Linearised Quantum Gravity in GAdS 4: Normal Coordinates' Representation and the Matter-Graviton Interaction Hamiltonian}}

In this section, we quantise gravity in $GADS_4$ and a one-dimensional (1D) quantum harmonic oscillator in Riemann Normal coordinates in AdS~\cite{BlauGR2025}, and then show that, in the limit $L\rightarrow \infty$, this reduces to the Fermi Normal coordinates~\cite{Rakhmanov2014}.

\subsection{Harmonic Oscillator in Global Anti-de Sitter Spacetime}

Let us consider the Hamiltonian of the 1D harmonic oscillator, which is given by
\begin{equation}
    H_m = \frac{p^2}{2m} + \frac{1}{2}m\omega_m^2x^2
    \label{CHO}
\end{equation}
where we assume $\omega_m \gg \omega=c/L$.
Consider the amplitude of the oscillator is $A$; we can write the oscillatory solution of the trajectory as $x = A\cos(\omega_m t)$. We can find the traceless quadrupole moment due to the harmonic potential in \eqref{CHO} by the non-zero components,
\begin{equation}
    D_{22} = D_{33} = -\frac{1}{2}D_{11} = - mA^2\cos^2(\omega_m t)
\end{equation}
where $x$ is the direction we denoted earlier as $x_1$. This harmonic oscillator will also source gravitational radiation. To be consistent, we need to quantise both matter and gravitons. The time-independent Schr\"odinger equation in $GAdS_4$ is given by,
\begin{align}
    H\psi(x) = \left[-\frac{\hbar^2}{2m}\Delta_g^s + V(x)\right]\psi(x) = E\psi(x);\nonumber \\ 
    \Delta_g\psi = \frac{1}{\sqrt{-g}}\partial_\mu(\sqrt{-g}g^{\mu\nu}\partial_\nu\psi)\,,
\end{align}
where $\Delta_g$ is the Laplace-Beltrami operator, and $\Delta_g^s$ is the spatial part of it. We expand the metric about the mean position of the oscillator by Riemann Normal Coordinates (RNC), given by \cite{BlauGR2025}
\begin{eqnarray}
    g_{00} &=& -1 - \frac{1}{3}R_{0 i 0 j}\, x^i x^j + \mathcal{O}(x^3)\,,\nonumber\\
    g_{ij} &=& \delta_{ij} - \frac{1}{3} R_{i k j l}\, x^k x^l + \mathcal{O}(x^3)
\end{eqnarray}
Writing the 3D Laplace-Beltrami operator in terms of these coordinates in the non-relativistic limit, we find, 
\begin{equation}
    \Delta_g = \nabla^2 + \frac{1}{3} R_{i k j l}\, x^k x^l \, \partial_i \partial_j - \frac{2}{3} R_{i j}\, x^j \partial_i + \mathcal{O}\!\left(x^2 \partial^2,\, x \partial\right)
\end{equation}
We now rescale the wavefunction by $\psi\longrightarrow \sqrt{-g}^{1/4}\psi$ to preserve the probability spectrum as in flat space, and noting that 
$$R_{ijkl} = -\frac{1}{L^2}(\delta_{ij}\delta_{kl} - \delta_{il}\delta_{jk})$$ and $$R_{0i0j} = -\frac{1}{L^2}\delta_{ij},$$ we find that,
\begin{eqnarray}
    H_{eff} &=& H_{flat} + \frac{\hbar^2}{2mL^2} + \frac{\hbar^2}{2m}\frac{R}{6} +\text{ }...\,,\nonumber \\ 
    H_{flat} &=& -\frac{\hbar^2}{2m}\nabla^2\,.
\end{eqnarray}
Thus, the geodesic deviation equation can be read off from this expansion of the non-relativistic (NR)-$GAdS_4$ Hamiltonian as $d^2x/dt^2 = - ({c^2}/{L^2})x\implies V_c = ({mc^2}/{2L^2})x^2$ (an effective potential). As noted before, the negative curvature induces an effective harmonic potential that is independent of the oscillator's own harmonic potential. Thus, up to second order, we obtain
\begin{equation}
    H_{eff} = \frac{\hat{p}^2}{2m} + \frac{1}{2}m \left(\omega_m^2 + \frac{c^2}{L^2}\right)x^2\,.
\end{equation}
We  define
\begin{equation}
\omega_{eff}^2 = \omega_m^2 + \frac{c^2}{L^2}\,.
\end{equation}
The effective length of the quantum harmonic oscillator is determined by the zero-point fluctuation scale, given by: 
\begin{equation}
\ell_{osc} = \delta_{zpf} = \sqrt{\frac{\hbar}{2m\omega_m}}
\label{zpf}
\end{equation}
Hence, it can clearly be seen that (we define $\delta\omega = \omega_{eff} - \omega_m$) the relative change of the oscillator frequency due to the intrinsic curvature of AdS spacetime is,
\begin{equation}
    \frac{\delta\omega}{\omega_m} = \frac{mc^2}{\hbar\omega_m}\left(\frac{\ell_{osc}}{L}\right)^2 = \frac{c^2}{2\omega_m^2 L^2} \ll 1
\end{equation}
as long as $\ell_{osc}\ll L$, which is clearly the physical limit. Here we have also used $\sqrt{1+x} \approx 1 + \frac{1}{2}x$. Hence, the effect of AdS space on the oscillator's inherent frequency can be neglected to second order, and the interaction term between the oscillator and the quantum spacetime can be approximated by the flat-space interaction derived from Fermi Normal Coordinates~\cite{Rakhmanov2014}. For a more complete review of quantum mechanics in curved spacetime, see \cite{ParkerToms2009}.


\subsection{Fermi Normal Coordinates and Linearised Quantum Gravity}

We now make a crucial assumption that will hold throughout the paper. We assume the matter is moving non-relativistically, meaning the oscillation speed is much lower than the speed of light, allowing us to separate the Hilbert spaces of the graviton and the oscillator. Thus, the coordinate system we choose to analyse the oscillator's motion can couple to the tidal force arising from the metric perturbation (which we will call a graviton) expressed in another coordinate system. This is also valid, as now we will work in Fermi Normal Coordinates - and everything is locally Cartesian and flat about the geodesic of the oscillator. But this will also imply that the interaction is local in spacetime, a fact we will observe in section IV. 

We will choose the local coordinates for the oscillator, $x^\mu = (-ct, x, y, z)$. We have, 
\begin{equation}
g_{\mu\nu} = \eta_{\mu\nu} + h_{\mu\nu}\,.
\end{equation}
The dominant contribution to the metric around the localized geodesic is given by (in the TT-gauge~\cite{Maggiore2008}, so that no remnant gauge freedom is remaining), see \cite{Rakhmanov2014}~\footnote{In general, $GAdS$ has Fermi-Normal coordinates for the $00$-component given by, $g_{00} = -1 -\frac{\rho^2}{L^2} - \frac{1}{2}\left.\frac{\partial^2h_{11}}{\partial t^2}\right|_{t=0}x^2$, for small $\rho\approx \sqrt{x^2 + y^2 + z^2} $ - but we have already showed that this acts like a tidal term anyway and we can ignore it as long as the oscillator is localized. So the action can be justified through this as well. Here, $g_{\mu\nu} = \bar{\eta}_{\mu\nu} + h_{\mu\nu}$, where, $\bar{\eta}$ is the metric of the background $GAdS$.}, 
\begin{eqnarray}
    g_{00} &=& -(1 + R_{0i0j}x^ix^j)\nonumber \\ 
    &=& -\left(1 + \frac{1}{2c^2}\left.\frac{\partial^2h_{11}}{\partial t^2}\right|_{t=0}x^2\right)\,.
    \label{FNC}
\end{eqnarray}
where, $\ddot{h}_{11}$, where dot denotes the derivative with respect to time $t$ is evaluated at the Fermi Normal coordinates. From Eq.~\eqref{PS} and the metric component written just above, we find (see Appendix~\ref{APPD} for more details),
\begin{equation}
    {H}_{int} = -L_{int} = \frac{m}{4}\left.\frac{\partial^2h_{11}}{\partial t^2}\right|_{t=0}x^2\,,
    \label{interaction}
\end{equation}
here we note that $H_{int},~h_{11},~x$ are all quantum operators. It can be seen that linearised quantum gravity behaves remarkably similarly to a massive free scalar field in the AdS spacetime\cite{Burgess2004}. Since canonical QFT is not Lorentz-covariant, to observe this effect we can move to a covariant phase-space formulation. The Witten-Crnkovic formulation of integrating a symplectic 2-form over a Cauchy surface in spacetime \cite{CrnkovicWitten1987} shows that the current is, up to the internal tensorial structure, equivalent to the Klein-Gordon Lorentz-covariant current for a scalar field in curved spacetime. Thus, if we take into account the exact polarisation tensor structure expected for an AdS graviton, the rest will behave exactly like the canonical scalar field in AdS spacetime. In other words, the Fierz-Pauli action~\cite{farolfi2025fierzpaulitheorycurvedspacetime, FierzPauli1939} up to the second order for massless gravitons in general curved spacetime can be given in the covariant TT-gauge as, 
\begin{equation}
(\Box h_{\mu\nu} - 2R_{\rho(\mu\nu)\lambda}h^{\rho\lambda}) = 0 \implies (\Box + \frac{2}{L^2})h_{\mu\nu} = 0.
\end{equation}
Thus, $m^2_{eff} = -\frac{2}{L^2} \implies \Delta = 3/2 + \sqrt{9/4 - 2} = 2$ 
\footnote{Note that, however, this is not equivalent to saying that gravity becomes massive in in AdS, as any metric perturbation in AdS - $g_{\mu\nu} = \bar{g}_{\mu\nu}+h_{\mu\nu}$, with $\bar{g}$ being the AdS metric - must satisfy a Fierz-Pauli \cite{FierzPauli1939} type of mass term uniquely to qualify as a massive term in the linearized gravity regime, such as, $\mathcal{L}_{\text{FP}} = m^2(h^{\mu\nu}h_{\mu\nu} - (\bar{g}^{\mu\nu}h_{\mu\nu})^2)$ to be a theory of a massive spin-2 field. Our equation of motion here in the linearised quantum regime does not imply any such term in the AdS action.}.

Thus, the graviton in $GAdS$ space is given by (with the dynamical degrees of freedom only coming from the spatial components; for details see section 7.2 of Ref.~\cite{Carroll:2004st}, and for the derivation, see Appendix \ref{SQ}),
\begin{eqnarray}
    h_{ij}(t,\rho,\Omega)
&=&
\sqrt{\frac{8\pi G\hbar}{c^2}}\sum_{n,\ell,m}
\frac{1}{\sqrt{\,c/L(2n+\ell+2)}}\nonumber\\
&\times& \Big[
a_{n\ell m\lambda}
\,e^{-ic/L(2n+\ell+2)t}
+
\text{H.c.}]
R_{n\ell}(\rho)\nonumber\\
&\times&\epsilon_{ij}^{(\ell,m,\lambda)}(\Omega); \omega_{n\ell} = c/L(2n+\ell+2)
\label{graviton equation}
\end{eqnarray}
where $a_{n\ell m}$ is the annihilation operator, and
\begin{eqnarray}
    R_{n\ell}(\rho) &=& \frac{N_{n\ell}}{\sqrt{\pi^2L^3}}\,
(\sinh\rho)^{\ell}\,
(\cosh\rho)^{-2}\times\,\nonumber\\
&{}&_2F_1\!\left(
 -n,\,
 2 + \ell + n;\,
 \ell + \tfrac{3}{2};\,
 \tanh^2\rho
\right)
\label{mq}
\end{eqnarray} 
and 
\begin{equation}
N_{n\ell}^2 = \frac{2\Gamma(n+1)\Gamma(n+\ell+3)}{\Gamma(n+\ell+3/2)\Gamma(n+5/2)}.
\end{equation}
For gravitons, $\Delta = 2$ in (1+3)-dimensional spacetime.  Note that 
$\ell \geq 2$ for the tensor spherical harmonics to be non-zero, whereas $n$ is a whole number and $m$ can run from $-\ell \leq m\leq +\ell$, thus having a degeneracy of 
$(2\ell + 1)$. 

We have implicitly assumed a summation over the polarisation modes. 
$\lambda = 1, 2$ in the polarisation tensor $\epsilon_{ij}^\lambda$. The corresponding kinetic term for the massless graviton field is,
\begin{equation}
    {H}_{\text{grav}} = \sum_{nlm} \, \hbar \omega_{nl} \, a^\dagger_{nlm,\lambda} \, a_{nlm,\lambda}
    \label{gravitational hamiltonian}
\end{equation}
In Eq.\eqref{CHO}, we introduce the dimensional amplitude quadrature, $$X = b + b^\dagger.$$ It is related to the position operator as $\hat{x} = \delta_{zpf}X$. Similarly, $$P = i(b - b^\dagger)$$ such that $\hat{p} = \sqrt{\frac{hm\omega_m}{2}}P$. Thus,
\begin{equation}
    {H}_m = \hbar\omega_m b^\dagger b
\end{equation}
Thus, the interaction Hamiltonian Eq.\eqref{interaction} can be rewritten as,
\begin{equation}
    {H}_{\text{int}} = \sum_{\lambda} \sum_{nlm} \, G_{nlm}^{\lambda} \, a_{nlm,\lambda} \, X^2 + \text{H.c.}
\end{equation}
where,
\begin{equation}
    G^\lambda_{nlm} = \sqrt{\frac{\pi G\hbar^3\omega_{nl}^3}{8c^2\omega_m^2}}\epsilon_{11}^{lm\lambda}R_{n\ell}(\rho)
    \label{coup}
\end{equation}
Hence, the total Hamiltonian is given by:
\begin{equation}
    {H}_{tot} = {H}_{\text{int}} + {H}_m + {H}_{grav}
\end{equation}
which describes the system with $\mathcal{H}_m\otimes\mathcal{H}_{grav}$ being the total Hilbert space of the cumulative system. Note that the interaction term has no explicit dependence on the properties of the system's matter content, except for the mass and the trapping frequency $\omega_m$, and, as such, the physics derived henceforth will be the same for the microscopic and mesoscopic systems. 


\section{\texorpdfstring{Gravitational decohereence}{Coherence due to Graviton Bath in GAdS 4}}\label{something}

Before we start our analysis. we should note that, even though we will take a Markovian approach to derive the master equation for the AdS graviton, this analysis will generally be valid only at zeroth order. This is because Markovian analysis does not consider the full picture when the bath itself has a nontrivial structure (in the case of AdS, the internal curvature term) and thus fails to account for the effects arising from the bath's structure. For a fuller picture, the analysis needs to be done in a completely non-Markovian way (which we leave as work for a future paper). Some ways of analyzing non-Markovian systems and deriving their master equations involve - Nakajima-Zwanzing projection technique \cite{ignatyuk2019masterequationopenquantum} (where instead of tracing out the Liouvillean part of the quantum mechanical system, the relevant part is projected out by a projection operator defined on the Hilbert subspace of the system), Redfield master equation \cite{Poole2012Redfield} (where the rotating-wave approximation is bypassed), and the path integral formulation of the influence functional \cite{FEYNMAN1963118} for open quantum systems. The influence functional 
is the most preferred in condensed matter literature \cite{Biryukov:2016vae} (from which the analysis of the current paper takes main inspiration) because of its manifest Lorentz-covariant approach in dealing with influence functionals - thus, straightforwardly extensible to the entanglement entropy formulation of field theory.

Even though our structural derivation will follow the Markovian approach closely, we will deviate sometimes in taking certain approximations over the von Neumann equation - thus, the best way to follow our analysis here will be the term "semi-Markovian". Specifically, because the AdS boundary is at a finite coordinate-time distance for massless trajectories, taking the integral limit to infinity is avoided in Eq.\eqref{eq:born_markov}, thereby giving rise to new physical insight of decoherence in the AdS background.

In this "discrete QFT" model of the gravitational field in AdS spacetime, we will obtain the dynamics of the matter system coupled to the graviton. We will assume that the graviton field Eq.~\eqref{graviton equation} is in a ground state, that is, in an initially empty bath,
\begin{eqnarray}
    \langle a^\dagger_{n\ell m\lambda}, a_{n'\ell'm'\lambda'}\rangle = 0
     \label{comm}\\
    \langle a_{n\ell m\lambda} a^\dagger_{n'\ell'm'\lambda'}\rangle = \delta_{nn'}\delta_{\ell\ell'}\delta_{mm'}
    \delta_{\lambda\lambda'} 
\end{eqnarray}
with, $\langle a^\dagger a^\dagger\rangle = \langle aa\rangle = 0$. 

We derive the quantum master equation for the matter system - by tracing out the gravitational field - a closed form of this generic derivation can be found in chapter 3.3 of \cite{BreuerPetruccione2002}. We denote the total statistical operator of the problem as $\rho^{tot}$ (the matter-wave system and the gravitational field), and by $\rho\text{ }(\rho^{(g)})$ the reduced statistical operator for the matter-wave system (the gravitational field), obtained by tracing away the gravitational field (the matter-wave system). The von Neumann equation can be expressed in the interaction picture as~\cite{BreuerPetruccione2002}
\begin{equation}
    \frac{d}{dt}\rho^{tot} = -\frac{i}{\hbar}[H_t^{int}, \rho_t^{tot}]
    \label{vNe}
\end{equation}
where the interaction Hamiltonian in the interaction picture is given by
\begin{equation}
   {H}_t^{int} = \sum_{\lambda} \sum_{nlm} \, G_{nlm}^{\lambda} \, a_{nlm,\lambda} \, X_t^2 + \text{H.c.}
   \label{inth}
\end{equation}
The interaction picture time-dependent amplitude quadrature is given by,
\begin{equation}
    X_t = be^{-i\omega_m t} + b^\dagger e^{i\omega_mt}
    \label{pos}
\end{equation}
The dynamics in Eq.~\eqref{vNe} can be formally solved by the following expression
\begin{equation}
    \rho^{tot}_t = \rho_0^{tot} - \frac{i}{\hbar}\int^t_0 ds[H^{int}_s, \rho_s^{tot}]
    \label{1vNe}
\end{equation}
Then, by inserting Eq.~\eqref{1vNe} into Eq.~\eqref{vNe} and tracing over the bath (which is equivalent to taking the vacuum expectation value with respect to the Bunch-Davies vacuum of the graviton field), we obtain,
\begin{equation}
    \frac{d}{dt}\rho^{tot}_t =\frac{1}{\hbar^2}\int^t_0 ds\ \text{tr}_g[H^{int}_t,[H^{int}_s,\rho^{tot}_s]]\,,
    \label{2vNe}
\end{equation}
where the first order term vanished as $\langle a\rangle = \langle a^\dagger\rangle = 0$. However, the second-order term on the RHS of Eq.~\eqref{2vNe} is nonzero because it depends on the normal-ordered vacuum fluctuations of the graviton field. Eq.~\eqref{2vNe} is an exact result, and it contains all the contributions from all the Feynman diagrams with any number of vertices. To derive an analogous Lindblad form of the quantum master equation - describing the effect of the dominant tree-level Feynman diagrams at the strength ${\cal O}(\sqrt{G})$ - we will need to perform certain approximations.

\begin{enumerate}

\item Born approximation: we assume the Born approximation first, $\rho^{tot}_s \approx \rho_s\otimes \rho^{(g)}$ on the RHS of Eq.~\eqref{2vNe}. It automatically precludes any entanglement between the matter-wave system and the gravitational field, as we are explicitly assuming a factorizable state to begin with. We also note that the absence of the subscript $s$ in the reduced gravitational field state is because the interaction between the matter-wave system and the spacetime is weak, and so this state changes only negligibly in the interaction picture. 

\item Semi-Markov approximation: We will now make Eq.~\eqref{2vNe} local in time such that the state of the equation only depends on $\rho_t$ at time $t$ and not on earlier times of the state. For this, we formally solve the von Neumann equation to connect the state at time $s$ with the state at time $t$ to find,
\begin{equation}
    \rho^{tot}_s = \rho^{tot}_t + \frac{i}{\hbar}\int_s^tds'[H^{int}_{s'}, \rho^{tot}_{s'}] 
    \label{3vNe}
\end{equation}
as we have done in Eq.~\eqref{1vNe}. We then insert Eq.~\eqref{3vNe} into Eq.~\eqref{vNe} to find,
\begin{align}
\frac{d\rho_t}{dt}
&=
-\frac{1}{\hbar^2}
\int_0^t ds \,
\mathrm{tr}_g
\left[
H^{\mathrm{int}}_t,
\left[
H^{\mathrm{int}}_s,
\rho^{\mathrm{tot}}_t
\right]
\right]
\nonumber \\
&\quad
-\frac{i}{\hbar^3}
\int_0^t ds
\int_s^t ds'
\,
\mathrm{tr}_g
\left[
H^{\mathrm{int}}_t,
\left[
H^{\mathrm{int}}_s,
\left[
H^{\mathrm{int}}_{s'},
\rho^{\mathrm{tot}}_{s'}
\right]
\right]
\right]
\nonumber \\
&\qquad\qquad
+ \mathcal{O}(G^2)
\label{eq:born_expansion}
\end{align}
The first term on the RHS of Eq.~\eqref{eq:born_expansion} is of order $G$, and the second is of order $\sqrt{G^3}$. We will keep only the dominant terms up to order $\sim G$ (each Hamiltonian operator induces a vertex of order $\sqrt{G}$). Since, the gravitational constant is sufficiently weak with respect to whatever the coupling constants may be governing the internal dynamics of the matter-wave system (pertaining to the standard model of interactions) this is well justified. Thus, Eq.~\eqref{eq:born_expansion} is completely local in time. We now change the integration variable $s\longrightarrow t-s$ to get,
\begin{equation}
\frac{d\rho_t}{dt}
=
-\frac{1}{\hbar^2}
\int_0^t ds \,
\mathrm{tr}_g
\left[
H_t^{\mathrm{int}},
\left[
H_{t-s}^{\mathrm{int}},
\rho_t^{\mathrm{tot}}
\right]
\right]
\label{eq:markov_master}
\end{equation}
Unlike flat space, we cannot employ Markov approximation as commonly expected in analogous electromagnetic calculations, see Ref.~\cite{GardinerZoller2004}, because we cannot a priori assume the AdS curvature to be so large that the non-relativistic matter (which our matter-wave system is) system's dynamics changes negligibly within the decay time of the graviton bath correlation function. Thus, the equation Eq.~\eqref{eq:markov_master} is sensitive to the choice of initial time - this will have important implications for the physics of the quantum master equation of a finite spacetime as we will shortly see.
\end{enumerate}

Therefore the final semi-Markovian master equation becomes:
\begin{equation}
\frac{d\rho_t}{dt}
=
-\frac{1}{\hbar^2}
\int_0^t ds \,
\mathrm{tr}_g
\left[
H_t^{\mathrm{int}},
\left[
H_{t-s}^{\mathrm{int}},
\rho_t \otimes \rho^{(g)}
\right]
\right]
\label{eq:born_markov}
\end{equation}
We then insert Eq.~\eqref{inth} in the Eq.~\eqref{eq:born_markov} to finally see,
\begin{widetext}
\begin{align}
\frac{d\rho_t}{dt}
=
-\frac{1}{\hbar^2}
\int_0^t ds
\sum_{\lambda,\lambda'}
\sum_{n\ell m}
\sum_{n'\ell'm'}
\langle
a_{n\ell m,\lambda}
a^\dagger_{n'\ell'm',\lambda'}
\rangle
G^\lambda_{n\ell m}
G^{\lambda'}_{n'\ell'm'}
e^{-i(\omega_{n\ell}-\omega_{n'\ell'})t}
\nonumber\\
\times
\Big(
e^{-i\omega_{n'\ell'}s}
X_t^2 X_{t-s}^2 \rho_t
-
e^{-i\omega_{n'\ell'}s}
X_t^2 \rho_t X_{t-s}^2
+
e^{i\omega_{n\ell}s}
\rho_t X_{t-s}^2 X_t^2
-
e^{-i\omega_{n\ell}s}
X_{t-s}^2 \rho_t X_t^2
\Big).
\label{eq:39}
\end{align}
\end{widetext}
We've already used the fact that there are no graviton excitations in the $GAdS$ spacetime initially (see Eq.~\eqref{comm}). We now insert the non-zero values of the vacuum fluctuations to get,
\begin{align}
\frac{d\rho_t}{dt}
=
-\frac{1}{\hbar^2}
\int_0^{t} ds
\sum_{n\ell m,\lambda}
\left(G^{\lambda}_{n\ell m}\right)^2
\Big(
e^{-i\omega_{n\ell}s}
X_t^2 X_{t-s}^2 \rho_t
\nonumber\\
-
e^{i\omega_{n\ell}s}
X_t^2 \rho_t X_{t-s}^2
+
e^{i\omega_{n\ell}s}
\rho_t X_{t-s}^2 X_t^2
-
e^{-i\omega_{n\ell}s}
X_{t-s}^2 \rho_t X_t^2
\Big)
\label{eq:ads_master_equation_simplified}
\end{align}
We now insert the coupling from Eq.~\eqref{coup} to finally find,
\begin{align}
\frac{d\rho_t}{dt}
= -\int_0^{t} ds
\sum_{n\ell m,\lambda}
\frac{
\pi G \hbar \omega_{n\ell}^{3}R^2_{n\ell}(\rho)
}{
8 c^{2} \omega_m^{2}
}
\,\epsilon^{n\ell m,\lambda}_{11}(\Omega)\,
\epsilon^{n\ell m,\lambda}_{11}(\Omega)
\nonumber \\
\times
\Big(
e^{-i\omega_{n\ell}s}
X_t^{2} X_{t-s}^{2} \rho_t
-
e^{i\omega_{n\ell}s}
X_t^{2} \rho_t X_{t-s}^{2}
\nonumber \\
\qquad\qquad
+
e^{i\omega_{n\ell}s}
\rho_t X_{t-s}^{2} X_t^{2}
-
e^{-i\omega_{n\ell}s}
X_{t-s}^{2} \rho_t X_t^{2}
\Big)
\label{eq:gads_master_equation}
\end{align}
We now note, 
\begin{equation}
\sum_{m, \lambda}\epsilon_{ij}^{(l,m,\lambda)}\epsilon_{kl}^{(l,m,\lambda)} = \frac{1}{4\pi}(2l+1)\Lambda_{ijkl}(\Omega),
\end{equation}
where
\begin{equation}
\Lambda_{ijkl} = \frac{1}{2}(P_{ik}P_{jl} + P_{il}P_{jk} - P_{ij}P_{kl})
\end{equation}
in the large-$L$ limit, tending to flat space. A slightly more general form of $\Lambda_{ijkl}$ is given in section \ref{reco}. We also note, 
\begin{equation}
P_{ij} = \delta_{ij} - \vec{n}_i\vec{n}_j
\end{equation}
where $\vec{n}$ is the transverse direction to the 2-sphere manifold with respect to the conserved 4-momentum of the gravitational waves coupled with the light-cone of the matter-wave system with respect to the gravitational field. By taking $i = j = k = l = 1$ in Eq.~\eqref{eq:gads_master_equation} we obtain
\begin{align}
    \frac{d\rho_t}{dt}
=
-
\int_0^{t} ds
\sum_{n\ell}
\frac{ G \hbar \omega_{n\ell}^{3}R^2_{n\ell}(\rho)
}{
64 c^{2} \omega_m^{2}
}(2\ell + 1)(\Lambda_{1111})^2
\nonumber \\
\times
\Big(
e^{-i\omega_{n\ell}s}
X_t^{2} X_{t-s}^{2} \rho_t
-
e^{i\omega_{n\ell}s}
X_t^{2} \rho_t X_{t-s}^{2}
\nonumber \\
\qquad\qquad
+
e^{i\omega_{n\ell}s}
\rho_t X_{t-s}^{2} X_t^{2}
-
e^{-i\omega_{n\ell}s}
X_{t-s}^{2} \rho_t X_t^{2}
\Big)
\label{int}
\end{align}
We now insert the position-amplitude observable from Eq.~\eqref{pos}. By defining 
\begin{equation}
N = bb^\dagger + b^\dagger b = 2b^\dagger b + 1\,,
\end{equation}
we will get the following terms,
\begin{align}
X_t^2 X_{t-s}^2
={}&\;
b^4 e^{-4i\omega_m t}e^{2i\omega_m s}
+b^2N\,e^{-2i\omega_m t}
+b^2b^{\dagger 2}e^{-2i\omega_m s}
\nonumber\\
&+Nb^2\,e^{-2i\omega_m t}e^{2i\omega_m s}
+N^2
+Nb^{\dagger 2}e^{2i\omega_m t}e^{-2i\omega_m s}
\nonumber\\
&+b^{\dagger 2}b^2e^{2i\omega_m s}
+b^{\dagger 2}N\,e^{2i\omega_m t}
+b^{\dagger 4}e^{4i\omega_m t}e^{-2i\omega_m s}
\end{align}
Similarly,
\begin{widetext}
\begin{align}
X_t^2 \rho_t X_{t-s}^2
={}&\;
b^2\rho_t b^2
e^{-4i\omega_m t}e^{2i\omega_m s}
+b^2\rho_t N
e^{-2i\omega_m t}
+b^2\rho_t b^{\dagger 2}
e^{-2i\omega_m s}
\nonumber\\
&+
N\rho_t b^2
e^{-2i\omega_m t}e^{2i\omega_m s}
+N\rho_t N
+N\rho_t b^{\dagger 2}
e^{2i\omega_m t}e^{-2i\omega_m s}
\nonumber\\
&+
b^{\dagger 2}\rho_t b^2
e^{2i\omega_m s}
+b^{\dagger 2}\rho_t N
e^{2i\omega_m t}
+b^{\dagger 2}\rho_t b^{\dagger 2}
e^{4i\omega_m t}e^{-2i\omega_m s}\,,
\label{eq:Xt2rhoXtms2}
\end{align}
\end{widetext}
and
\begin{widetext}
\begin{align}
\rho_t X_{t-s}^2 X_t^2
={}&\;
\rho_t b^4
e^{-4i\omega_m t}e^{2i\omega_m s}
+\rho_t N b^2
e^{-2i\omega_m t}e^{2i\omega_m s}
+\rho_t b^{\dagger 2}b^2
e^{2i\omega_m s}
\nonumber\\
&+
\rho_t b^2N
e^{-2i\omega_m t}
+\rho_t N^2
+\rho_t b^{\dagger 2}N
e^{2i\omega_m t}
\nonumber\\
&+
\rho_t b^2b^{\dagger 2}
e^{-2i\omega_m s}
+\rho_t Nb^{\dagger 2}
e^{2i\omega_m t}e^{-2i\omega_m s}
+\rho_t b^{\dagger 4}
e^{4i\omega_m t}e^{-2i\omega_m s}\,,
\end{align}
\end{widetext}
and
\begin{widetext}
\begin{align}
X_{t-s}^2 \rho_t X_t^2
={}&\;
b^2\rho_t b^2
e^{-4i\omega_m t}e^{2i\omega_m s}
+b^2\rho_t N
e^{-2i\omega_m t}e^{2i\omega_m s}
+b^2\rho_t b^{\dagger 2}
e^{2i\omega_m s}
\nonumber\\
&+
N\rho_t b^2
e^{-2i\omega_m t}
+N\rho_t N
+N\rho_t b^{\dagger 2}
e^{2i\omega_m t}
\nonumber\\
&+
b^{\dagger 2}\rho_t b^2
e^{-2i\omega_m s}
+b^{\dagger 2}\rho_t N
e^{2i\omega_m t}e^{-2i\omega_m s}
+b^{\dagger 2}\rho_t b^{\dagger 2}
e^{4i\omega_m t}e^{-2i\omega_m s}\,.
\end{align}
\end{widetext}
Since we are working in the interaction picture, also, we assume that $\rho_t$ evolves slowly compared to the harmonic oscillator evolution. Any terms, therefore, that will retain $e^{\pm 4i\omega_m t}$ will oscillate rapidly and average to zero over long times. We will apply the rotating-wave approximation~\cite{BreuerPetruccione2002,Agarwal2012}; that is, we keep only terms with equal numbers of $b$ and $b^\dagger$ (which retain the secular protection of the graviton-harmonic oscillator interaction) and neglect the other fast-rotating term (see Ref.~\cite{Agarwal2012}).

To make sure that only the nearly secular-term (equivalently, nearly-resonant term - $\omega_{n\ell}\approx 2\omega_m$), survives, 
for example, we take $e^{-i\omega_{n\ell}s} X_t^{2} X_{t-s}^{2} \rho_t$ - we need a $e^{2i\omega_ms}$ phase from the position operators. Multiplying by the bath phase gives, $e^{-i\omega_{n\ell}s} (b^{\dagger 2} b^2 e^{2i\omega_m s}) \rho_t = e^{-i(\omega_{n\ell} - 2\omega_m)s} b^{\dagger 2} b^2 \rho_t$ - which is a coveted term. We finally find,
\begin{align}
\frac{d\rho_t}{dt}
=
- \sum_{n\ell}(2\ell + 1)R^2_{n\ell}(\rho)
\frac{
G \hbar \omega_{n\ell}^{3}
}{
64 c^{2}\omega_m^{2}
}(\Lambda_{1111})^2\nonumber\\
\int_0^{t} ds
\Big(
e^{-i(\omega_{n\ell}-2\omega_m)s}
\, b^{\dagger 2} b^{2}\rho_t
-
e^{-i(\omega_{n\ell}-2\omega_m)s}
\, b^{2}\rho_t b^{\dagger 2} \nonumber\\
+
e^{i(\omega_{n\ell}-2\omega_m)s}
\, \rho_t b^{\dagger 2} b^{2}
-
e^{i(\omega_{n\ell}-2\omega_m)s}
\, b^{2}\rho_t b^{\dagger 2}
\Big)
\label{eq:gravitational_master_equation}
\end{align}
We simplify to find,
\begin{align}
\frac{d\rho_t}{dt}
&=
-\sum_{n\ell}
(2\ell+1)
\frac{
G\hbar\omega_{n\ell}^{3}
}{
64c^{2}\omega_m^{2}
}
R_{n\ell}^{2}(\rho)(\Lambda_{1111})^2
\nonumber \\
&\int_0^t ds
\Big[
e^{-i(\omega_{n\ell}-2\omega_m)s}
\left(
b^{\dagger 2} b^{2}\rho_t
-
b^{2}\rho_t b^{\dagger 2}
\right)
\nonumber\\
&+
e^{i(\omega_{n\ell}-2\omega_m)s}
\left(
\rho_t b^{\dagger 2} b^{2}
-
b^{2}\rho_t b^{\dagger 2}
\right)
\Big]
\label{eq:gads_master_equation_discrete}
\end{align}
We can evaluate the $s$-integral to simplify Eq.~\eqref{eq:gads_master_equation_discrete} as,
\begin{align}
    \frac{d\rho_t}{dt}
&=
-\sum_{n\ell}
(2\ell+1)
\frac{
G\hbar\omega_{n\ell}^{3}
}{
64c^{2}\omega_m^{2}
}
R_{n\ell}^{2}(\rho)(\Lambda_{1111})^2\times\nonumber\\
&\Big[
\alpha_R
\left(
\{O,\rho_t\}
-
2 b^{2}\rho_t b^{2}
\right)
-
i\alpha_I
\left[
b^{2}b^{2},
\rho_t
\right]
\Big]
\label{eq:lindblad_ads}
\end{align}
where, 
\begin{equation}
O = b^{\dagger2}b^2,~~\alpha_R = \frac{\sin(\Delta\omega t)}{\Delta\omega},~~ \alpha_I = \frac{1 -\cos(\Delta\omega t)}{\Delta\omega},
\end{equation}
and 
\begin{equation}
\Delta\omega = \omega_{n\ell} - 2\omega_m
\label{momentum conservation}
\end{equation}
We have used the usual notation $\{.\}$ for the anti-commutator and $[.]$ for the commutator of quantum operators. 

Before discussing the consequences of the final master-modified Lindblad equation in Eq.~\eqref{eq:lindblad_ads}, we note that the equation is valid over a wide range of particle masses, from the microscopic to the mesoscopic scale and beyond, as long as the Compton wavelengths of these masses are less than the AdS length scale. 

\subsection{Decoherence rate and selection rules}
In the last section, we derived the quantum master equation Eq.~\eqref{eq:lindblad_ads} dynamics of a harmonically trapped particle in a "discrete QFT" model of a graviton bath in $GAdS_4$. We show that the $GAdS_4$ master equation recovers the flat-space equation in appropriate limits in Appendix~\ref{flat}, and discuss the vacuum-energy shift of the reduced density matrix in Appendix~\ref{Lamb}. In this section, we derive certain selection rules of the AdS graviton modes based on the dissipation term, and derive the decoherence/recoherence time scale based on the same.

We define the decoherence coefficient from Eq.~\eqref{eq:lindblad_ads}, which quantifies the dynamics contribution of the mode $(n, \ell)$, summing over all of which will yield the net evolution of the density matrix coupled with the Lindblad dissipator; we define the decoherence rate of the quantum harmonic oscillator in a $GAdS_4$ as:
%
\begin{equation}
    \Gamma_{n\ell}(\rho, t) = (2\ell + 1)\frac{G\hbar\omega_{n\ell}^3}{64c^2\omega_m^2}R^2_{n\ell}(\rho)(\Lambda_{1111})^2\frac{\sin(\Delta\omega t)}{\Delta\omega}
    \label{Dissipation}
\end{equation}
From Eq.~\eqref{Dissipation} we see that the dissipation rate of the oscillation function is given by $\frac{\sin(\Delta\omega t)}{\Delta\omega}$\footnote{Under the flat-space limit, $\Delta\omega = 0$, so, $\frac{\sin(\Delta\omega t)}{\Delta\omega}$ goes to 1 at that limit. Therefore, the rest of the modes turn into a spatially dependent contribution to the decoherence, which, under proper limits and summed over all possible modes (with the summation being replaced by an integral), as shown in \ref{flat}, reduces to the constant flat-space decoherence coefficient - $\gamma_{\text{grav}} = \frac{32}{15}t^2_{\text{Pl}}\omega_m^3$.}. The physical time window on which this function turns negative is the exact interval on which recoherence happens - and this revival time must be periodic in $2\pi I,~~ I\in (0, 1, 2, ...)$ and in principle, be equal to integer multiples of the time-scale calculated in Eq.~\eqref{time}.

We define the revival/recoherence time as the periodic intervals upon which the initial full coherence will be regained if we denote the starting time of our analysis as $t=0$. The first revival time is therefore given by
\begin{equation}
    t_1 = \frac{2\pi}{c/L(2n+\ell + 2) - 2\omega_m} = \frac{\pi L}{2c}
\end{equation}
Simplifying, we get 
\begin{equation}
    \frac{c}{L}(2n + \ell + 2) = \omega_{n\ell} = \frac{4c}{L} + 2\omega_m
    \label{s1}
\end{equation}
This is the selection rule for graviton-mode frequencies, as implied by the semi-Markovian master equation derived in Eq.~\eqref{eq:lindblad_ads}. We can rewrite Eq.~\eqref{s1} as, 
\begin{equation}
    2n + \ell = 2\left(1+\frac{\omega_m L}{c}\right); \ell \geq 2
    \label{selection rules}
\end{equation}
The flat-space limit can be easily checked using Eq.~\eqref{s1}. Revoking the flat-space conditions, $n\longrightarrow\infty, \ell \sim \text{finite}, L \longrightarrow\infty$, we can recast both the LHS and the RHS of \eqref{s1} as $2cn/L \sim \omega_k \longrightarrow 2\omega_m$ - thus reviving the flat-space selection rule, $\omega_m = \omega_k/2$. 
\\Thus, there are effectively two selection rules:-
\begin{itemize}
    \item Local momentum conservation: Eq.~\eqref{momentum conservation} is the local quantification of momentum transfer between the harmonic oscillator and gravitons - the vertex contribution. Thus, in the limit that $L\longrightarrow\infty$ (the flat-space limit), which is equivalent to taking the limit of $t$ to $\infty$ in \eqref{int}, \eqref{momentum conservation} reduces to a sinc function; thus $\Delta\omega = 0\implies \omega_k = 2\omega_m$, which is precisely the flat-space momentum conservation equation, see~\cite{Toros:2020krn}.
    
    \item Global: Eq.~\eqref{selection rules} is the global conservation rule, which comes from matching the revival time with the light-crossing time in AdS - the latter is a property of the large-scale structure of the AdS spacetime. Therefore, $\omega_m L/c$ can be either an integer or a half-integer for Eq.~\eqref{selection rules} to be satisfied.
\end{itemize}
It seems that - much like the non-relativistic quantum mechanical eigenvalue equation for the Hamiltonian of the hydrogen atom - the structure of the equation implies that the harmonic oscillator frequency dictates which curvatures are allowed for the trapping to be valid for the background AdS spacetime, and vice-versa. However, in general, this is not the case, as the harmonic oscillator is a local system and its physics cannot affect the global calculations. Indeed, we saw from Eq.~\eqref{momentum conservation} that the tree-level calculation of the QFT model implied a local momentum-conservation rule. Therefore, we can interpret Eq.~\eqref{s1} as a modified selection rule for the AdS graviton frequency - on satisfaction of which there will be a global recoherence/revival of the dissipation coefficient. Therefore, for the specific combinations of $\omega_m$ and $L$ mentioned above, information will never be completely lost (equivalently, will never decohere). 


\subsection{Example of recoherence}\label{reco}

In this subsection, we estimate the recoherence time scale for specific quantum numbers.

For demonstration purposes, we take an example $\frac{\omega_m L}{c} = 2$,
then, $(n, \ell) = (2, 2), (1, 4), (0, 6)$, from \eqref{selection rules} - these are three possible combinations allowed from this particular value of $\omega_m$ and each of the modes implies that the coherence coefficient becomes negative at the same time for the whole master equation. Therefore, all the individual modes will have the same physics. For brevity, we explicitly calculate the dissipation coefficient associated with the $(2, 2)$-mode. 

We consider the harmonic oscillator to be localised around $\rho \sim 0$, or, in other words, $\rho$ is very small. For calculating a form of $\Lambda_{1111}$, we make a clever approximation - we assume $L$ is large enough that we will be able to attach a local Cartesian frame to the harmonic oscillator. In such case, $\epsilon_{ij}^{\ell m\lambda}$ is transverse-traceless with respect to $$\gamma_{ab}dx^adx^b = d\theta^2 + \sin^2\theta d\phi^2.$$ We take a unit vector in $GAdS_4$ - $(\hat{\rho}, \hat{\theta}, \hat{\phi})$. Let, the oscillator is oscillating along the $\hat{z}$ in the local Cartesian frame attached to it (aligned along the $1$-direction in the GAdS frame) - thus, $\hat{z} = \cos\theta\hat{\rho} - \sin\theta\hat{\theta}$. 
Because the polarisation tensors are transverse to the radial direction, $\epsilon_{\rho\mu} = 0$. Thus, 
\begin{equation}
    \epsilon_{11} = \epsilon_{zz} = \epsilon_{\hat{\theta}\hat{\theta}}(\hat{z}\cdot\hat{\theta})^2 + \epsilon_{\hat{\phi}\hat{\phi}}(\hat{z}\cdot\hat{\phi})^2 + 2\epsilon_{\hat{\theta}\hat{\phi}}(\hat{z}\cdot\hat{\theta})(\hat{z}\cdot\hat{\phi})
\end{equation}
Since, $\hat{z}.\hat{\phi} = 0$, we simplify the above expression,
\begin{equation}
    \epsilon_{11} = \epsilon_{\hat{\theta}\hat{\theta}} \sin^2\theta
\end{equation}
The TT-components for the $+$ and $\times$ polarizations can be physically represented as weighted sum over spin-2 spherical harmonics,
\begin{eqnarray}
    \epsilon_{\hat{\theta}\hat{\theta}}^+ = -\epsilon_{\hat{\phi}\hat{\phi}}^+ &=& \frac{1}{2\sqrt{2}} \left( {}_{+2}Y_{lm} + {}_{-2}Y_{lm} \right)\nonumber\\
    \epsilon_{\hat{\theta}\hat{\phi}}^\times = \epsilon_{\hat{\phi}\hat{\theta}}^\times &=& \frac{1}{2\sqrt{2}i} \left( {}_{+2}Y_{lm} - {}_{-2}Y_{lm} \right)
\end{eqnarray}
Thus, 
\begin{equation}
    \sum_{\lambda} |\epsilon_{\hat{\theta}\hat{\theta}}^\lambda|^2 = \frac{1}{8}\left|{}_{+2}Y_{lm} + {}_{-2}Y_{lm}\right|^2
\end{equation}
We now sum over all possible $m$ for a given $\ell$. Thus, we get,
\begin{align}
    \sum_{m=-l}^l |\epsilon_{\hat{\theta}\hat{\theta}}^+|^2 = \frac{1}{8}\sum_{m=-l}^l ( |{}_{+2}Y_{lm}|^2 + |{}_{-2}Y_{lm}|^2\nonumber\\
    + 2\text{Re}\left({}_{+2}Y_{lm} {}_{-2}Y_{lm}^*\right))
\end{align}
The Wigner d-function representation of the spin-weighted spherical harmonics is given by
\begin{equation}
    {}_sY_{lm}(\theta, \phi) = \sqrt{\frac{2l+1}{4\pi}} d^l_{-s, m}(\theta) e^{im\phi}
\end{equation}
These imply, 
\begin{equation}
    \sum_{m=-l}^l {}_{+2}Y_{lm} {}_{-2}Y_{lm}^* \propto \sum_{m=-l}^l d^l_{2,m}(\theta) d^l_{-2,m}(\theta) = \delta_{2, -2} = 0
\end{equation}
We now note, 
\begin{equation}
    \sum_{m=-l}^l \left| {}_2Y_{lm}(\theta, \phi) \right|^2 = \frac{2l+1}{4\pi}
\end{equation}
Altogether, we get,
\begin{equation}
    \sum_{m=-l}^l \sum_{\lambda} |\epsilon_{\hat{\theta}\hat{\theta}}^\lambda|^2 = \frac{2l+1}{16\pi} = \frac{2l+1}{4\pi}\left(\frac{1}{4}\right)
\end{equation}
Thus, by comparing with \eqref{eq:gads_master_equation}, we can say,
\begin{equation}
    \Lambda_{1111}(\theta, \phi) = \frac{1}{4}\sin^4\theta
    \label{lambda}
\end{equation}
Since $GAdS_4$ is an isotropic spacetime and there is a $\phi$-like isometry, we can set $\theta = \pi/2$ to get $\sin\theta = 1$. Thus, 
\begin{equation}
    (\Lambda_{1111})^2 = \frac{1}{16}
\end{equation}
Now, in our case, $\omega_m = 2c/L$, $\omega_{22} = 8c/L$, and therefore, $\Delta\omega = 6c/L$. Now, note that the hypergeometric function part of $R_{22}(\rho)$ has the 
2-1 Gaussian structure \cite{nicholson2019quadratictransformationshypergeometricfunction}, and hence, the polynomial representation terminates after degree $a$,
\begin{equation}
    {}_2F_1(a, b; c; z) = 1 + \frac{ab}{c}z + \frac{a(a+1)b(b+1)}{c(c+1)2!}z^2 + \dots
\end{equation}
In our case, this becomes a quadratic polynomial. Therefore, 
$${}_2F_1(-2, 6; 7/2; \tanh^2\rho) \implies R^2_{22} \approx \frac{N^2_{22}}{\pi^2 L^3} \rho^4$$ (for small $\rho$). Noting this, and evaluating, $N^2_{22} \approx \frac{234.38}{\pi}$, we find, altogether,
\begin{align}
    \Gamma_{22}(\rho, t) = 36.622\frac{G\hbar\rho^4}{\pi^3 L^3 c^2}\sin(\frac{4ct}{L}) \nonumber\\
    = 36.622\frac{t^2_{\text{Pl}}\omega_m^3\rho^4}{\pi^3}\sin(\frac{4ct}{L})
\end{align}
where we have used $\omega_m=2L/c$. 

Thus, the AdS physics enters into the master equation's dissipation coefficient through the embedding curvature radius of the spacetime. 

Similarly, for the other two modes, we get,
\begin{eqnarray}
    \Gamma_{14}(\rho, t) &=& 9.309\frac{t^2_{\text{Pl}}\omega_m^3\rho^8}{\pi^3}\sin(\frac{4ct}{L})\\
    \Gamma_{06}(\rho, t) &=& 41.373 \frac{t^2_{\text{Pl}}\omega_m^3\rho^{12}}{\pi^3}\sin(\frac{4ct}{L})
\end{eqnarray}
Here, we defined the Planck time as, $t_{\text{Pl}} = \sqrt{G\hbar/c^5} \sim 10^{-43}s$.
Thus, we can clearly see that for a given value of $\omega_m L/c$, the mode with the lowest possible $\ell$ will be the leading contribution to the coherence coefficient. This is because as $\ell$ increases, the dependence of $\Gamma_{n\ell}$ on $\rho$ increases in exponents of $\rho$, and since $\rho = \sinh^{-1}(r/L) \sim r/L$ for small $\rho$ the contribution will be lesser and lesser dominating. The numerical coefficient will always come out to be around the same order. 

\section{Conclusion}
In this paper, we took a previously unexplored approach of understanding how quantum mechanics (through its innate informational structure) can help us understand how a UV-complete quantum theory of gravity will look like in AdS spacetime. After demonstrating the structure and some key aspects of GAdS spacetime in hyperbolic coordinates, we first took a massive free particle in AdS spacetime and showed that the negativity of cosmological constant induces an attractive harmonic potential on the particle. Thus, it gains a frequency which is dependent on the UV-physics of the AdS spacetime. We showed that when we took a quantum harmonic oscillator, and treated the mechanics in classical AdS spacetime, this precise frequency comes as a correction to the energy eigenvalues. We argued why this correction's global nature will help us to approximate the system with high precision with Fermi normal coordinates rather than the more natural choice of Riemann natural coordinates in curved spacetime. 

We then derived the interaction Hamiltonian between the gravitational field and the quantum harmonic oscillator as they are assumed entangled - and then quantized gravity to leading order (in case of gravitons, it is 2nd order) using the Fierz-Pauli action being symplectically equivalent to the massive scalar field action in AdS. 

We then argued why the system is in principle a non-Markovian system, but the harmonic oscillator being a local system with no knowledge of the global structure of the spacetime, a carefully modified Markovian approach in understanding the entanglement structure of the system is sufficient to hint us towards the new physics that is expected of AdS.
\\Then, using a semi-Markovian analysis (where, to respect the non-trivial boundary structure of AdS spacetime we modified the upper limits of the time-integral in Markovian approximation, alongside the usual steps) we found out the von Neumann equation to the leading order of perturbation, and showed that the dynamics of the density matrix of the system hints that for certain choices of frequencies of the graviton the system actually never loses its information globally. Thus, the system does not decohere completely; it recoheres. We showed that this dynamics is not just a makeshift construct of the AdS spacetime, but actually reduces to the flat-space structure of the analogous equation under appropriate limits. 


Moreover, unlike the Casimir energy as expected in the flat spacetime master equation, we find that in Eq.~\eqref{eq:lindblad_ads}, the vacuum energy is Lamb-shifted finitely in the AdS master equation, much like the analogous system in cavity QED.

The flat-space limit of massless particles in AdS spacetime does not lead to plane waves, includes several energies and is dependent on the conformal dimension of the underlying field, as expected from states prepared from conformal primary wave functions in celestial holography \cite{Berenstein:2025tts}. 

A complete quantum theory of gravity is expected to exhibit UV/IR mixing, so IR parameters need not flow smoothly to their UV-complete counterparts \cite{agmon2023lecturesstringlandscapeswampland}. Holographic covariant entropy bounds \cite{Bousso_1999} indicate how the UV/IR mixing constrains EFTs within the Swampland program; just like dynamical horizon evolves in dS spacetime through UV/IR mixing\cite{cribiori2026cosmologicalconstraintsuvirmixing}, we can speculate that Eq.~\eqref{selection rules} suggests a holographically complete quantum theory of gravity in AdS constrains the cosmological constant (the UV parameter) through the local harmonic oscillator frequency (the IR parameter), thereby restricting the allowed AdS graviton spectrum.

\section{Acknowledgments}

M.B. would like to acknowledge the Institute fellowship awarded by IACS, Kolkata, part of which supported the research reported in this paper. M.B. would also acknowledge helpful discussions with Dr. Sumanta Chakraborty about the approach to the problem, and with Mr. Sounav Sengupta, who helped him realise the true potential of computational calculations while working on this paper. A.M.’s research is funded by the Gordon and Betty Moore Foundation through Grant GBMF12328, DOI 10.37807/GBMF12328. This material is based on work supported by the Alfred P. Sloan Foundation under Grant No. G-2023-21130

\appendix

\section{\texorpdfstring{Scalar Quantization in $GAdS_4$}{Scalar Quantization in GAdS 4}}\label{SQ}

The massive free scalar action in a general curved spacetime is given by,
\begin{equation}
    S = -\int d^4x\sqrt{-g}(\frac{1}{2}\nabla_\mu\phi\nabla^\mu\phi + m^2\phi^2)\,.
\end{equation}
The resulting equation of motion becomes $(\Box - m^2)\phi = 0$. Explicitly, for GAdS metric Eq.~\eqref{AdSm}, we have,
    \begin{eqnarray}
       \left[
-\frac{1}{\cosh^2 \rho}\,\partial_t^2
+ \frac{1}{\sinh^{2}\rho}\,\partial_\rho\!\left( \sinh^{d-1}\rho \,\partial_\rho \right)\right.\nonumber\\
+\left. \frac{1}{\sinh^2 \rho}\,\nabla_{S^{2}}^2
- m^2 L^2\right]\phi = 0
    \end{eqnarray}
    We use the separation of variables, such that, $$\phi(t,\rho,\Omega) = e^{-i\omega t}\, Y_{\ell m}(\Omega)\,R_{\omega\ell}(\rho),$$ where, $$\nabla_{S^{d-1}}^2\, Y_{\ell m}(\Omega) = -\ell(\ell + 1)\, Y_{\ell m}(\Omega).$$ 
We define the variable, $\Delta_{\pm} = \frac{3}{2} \pm \sqrt{\frac{9}{4} + m^2L^2}$. Clearly, the BF bound\footnote{The Breitenlohner-Freedman bound is a bound on the stability of quantum fields in (d+1)-AdS spacetime, which is given by, $m^2L^2 \geq -\frac{d^2}{4}$. If this is satisfied, then the mass term of the quantum field effectively produces no ghost-like kinetic term at any order of perturbative expansion of the scattering matrix.}, that is, the term under the square root of these variables, must be greater than zero. The Dirichlet boundary condition that we can impose now, as $\rho\longrightarrow\infty$, gives us the asymptotic behaviour of the field, $$\phi(\rho) \sim A\, e^{-\Delta_- \rho} + B\, e^{-\Delta_+ \rho}.$$ 
If we impose $A = 0$ (standard quantisation), then it imposes a reflective boundary condition at the boundary\footnote{If the BF bound is satisfied, then an alternate quantisation can be imposed such that the boundary becomes effectively spacelike, $B=0$.}. 
The radial equation becomes, 
\begin{align}
    \frac{1}{\sinh^{2}\rho}\,
\partial_\rho\!\left(
\sinh^{2}\rho\, \partial_\rho R
\right)
- \frac{\ell(\ell+1)}{\sinh^2\rho}\, R\nonumber\\
+ \frac{\omega^2}{\cosh^2\rho}\, R
- m^2 L^2\, R
= 0\,.
\label{EDM}
\end{align}
At $\rho\longrightarrow0$, for regularity, we choose, $R(\rho)\sim\rho^\ell$. We can define the inner product on this field, 
\begin{widetext}
\begin{equation}
\langle \phi_1 , \phi_2 \rangle = i \int d\rho \, d\Omega_{d-1}\, L^{d-1}(\sinh\rho)^{d-1} \cosh\rho\, \left(\phi_1\,\partial_t \phi_2 - \phi_2\,\partial_t \phi_1 \right).
\end{equation}
\end{widetext}
Therefore, the choice of $B=0$ is necessary to ensure a finite Klein-Gordon norm; other choices give rise to non-normalizable modes.  
These form an orthonormal Sturm-Liouville set of radial solutions, given by,
\begin{widetext}
\begin{align}
    R_{n\ell}(\rho)
= N_{n\ell}\,
(\sinh\rho)^{\ell}\,
(\cosh\rho)^{-\Delta}\,
{}_2F_1\!(
 -n,\,
 \Delta + \ell + n;
 \ell + \tfrac{3}{2};\,
 \tanh^2\rho)
\end{align}
\end{widetext}
where $_2F_1$ is the Gauss hypergeometric function of the 2-1 kind \cite{Andrews1999}. We impose the orthonormality condition: 
\begin{eqnarray}
\langle \phi_{n\ell m}, \phi_{n'\ell' m'} \rangle
= \delta_{n n'}\, \delta_{\ell \ell'}\, \delta_{m m'} \nonumber\\
\implies \int_{0}^{\infty} d\rho \, (\sinh \rho)^2 \, 
R_{n\ell}(\rho)\, R_{n'\ell'}(\rho)
= \delta_{n n'}\delta_{\ell\ell'}\,.
\end{eqnarray}
We can use this completeness to find the expression of the normalisation constant, which is written after Eq.~\eqref{mq}. 

The discrete spectrum turns out to be $\omega_{nl} = 2n+l+\Delta_+$, which is also the mass gap of the spectrum, which can be easily checked by substituting the solution back into Eq.~\eqref{EDM}.
If we define the canonical momentum with respect to the global timelike Killing vector $\partial_t$, we have $\Pi = \frac{\partial L}{\partial(\partial_t\phi)} = \sqrt{-g}g^{tt}\partial_t\phi$. Then, we impose the standard Dirac equal-time commutator quantisation, with $\hbar = 1$, $$[\phi(t, \vec{x}), \Pi(t, \vec{x'})] = i\delta^{(d)}(\vec{x} - \vec{x'}).$$
\\That implies, 
$$\left[
a_{n\ell m},
a_{n'\ell' m'}^{\dagger}
\right]
\propto \delta_{n n'}\, \delta_{\ell \ell'}\, \delta_{m m'}.$$ 
Therefore, we get,
\begin{eqnarray}
    \phi(t,\rho,\Omega)
= \sum_{n,\ell,m}
\left[
a_{n\ell m}\, u_{n\ell m}(t,\rho,\Omega)
+ a_{n\ell m}^{\dagger}\, u_{n\ell m}(t,\rho,\Omega)
\right]\nonumber\\
u_{n\ell m}(t,\rho,\Omega)\nonumber\,,
\end{eqnarray}
where, $u_{n\ell m}(t,\rho,\Omega)
= \frac{1}{\sqrt{2\omega_{n\ell}}}\,
e^{-i\omega_{n\ell} t}\,
Y_{\ell m}(\Omega)\,
R_{n\ell}(\rho)$. 
So, finally, we have,
\begin{align}
    \phi(t,\rho,\Omega)
= \sum_{n=0}^{\infty}
\sum_{\ell=0}^{\infty}
\sum_{m}
\frac{1}{\sqrt{2(2n+\ell+\Delta)}}\nonumber\\
[
a_{n\ell m}\, e^{-i(2n+\ell+\Delta)t}
+
\text{H.c.}]
Y_{\ell m}(\Omega)\,
R_{n\ell}(\rho)\,.
\label{sq1}
\end{align}
Note that $m$ here is a vector-valued magnetic quantum number, and the summation runs from $|\vec{m}| = -l\text{ to }+l$. $n$ and $\ell$ are independent of each other, and $n, \ell = 0, 1, 2, ...$. 
The vacuum here is Bunch-Davies~\cite{ParkerToms2009}, $$a_{nlm}\ket{0} = \ket{0}.$$ 
For check, the expansion in \eqref{sq1} should be divided by the effective volume experienced by the scalar modes in the expansion. For that, we will calculate the density of states in the AdS spacetime for this quantum scalar field and compare it with the density of states for quantum scalar fields in a finite volume, 
\begin{equation}
    g(\omega) = \frac{V\omega^2}{2\pi^2c^3}\,.
    \label{scalar density}
\end{equation}
We denote, $2n+\ell = q\implies \omega_{n\ell} = {q+\Delta}$. We will take $c = L = 1$ for this derivation. We want to find $\omega_{n\ell}\leq\omega\implies q\leq Q$, where $Q = \omega - \Delta$. We take the large-Q limit, hence, $\Delta \sim 0$. Since, $\ell$ has degeneracy $(2\ell + 1)$, we will have,
\begin{equation}
    N(Q) = \sum_{\ell=0}^Q(2\ell+1)\sum_{n=0}^{(Q-\ell)/2}(1) = \frac{1}{2}\sum_{\ell = 0}^Q(2\ell + 1)(Q-\ell +2)\,.
\end{equation}
Using the two well-known identities, $\sum_{\ell = 0}^Q\ell^2 = \frac{Q(Q+1)(2Q+1)}{6}$ and $\sum_{\ell = 0}^Q\ell = \frac{Q(Q+1)}{2}$, we will get by a straightforward and tedious calculation,
\begin{equation}
    N(Q) = \frac{1}{2}[(Q+2)^2Q - \frac{Q(Q+1)(4Q+5)}{6}]
\end{equation}
In the the large-Q limit, $N(Q)\approx Q^3/6$. But, we know, $Q = \omega\implies N(\omega) = \omega^3/6$. Thus,
\begin{equation}
    g(\omega) = \frac{dN}{d\omega} = \frac{\omega^2}{2}\implies g(\omega) = \frac{L^3\omega^2}{2c^3}\,.
    \label{DOS}
\end{equation}
Comparing with Eq.~\eqref{scalar density}, we can see the effective volume of the AdS spacetime to be $V_{eff} = \pi^2L^3$. Thus, scalar particles - and for that matter, any integer spin bosons, including gravitons - behave as if they are trapped in a gravitational box of dimension $\sim L$ in the AdS spacetime. 
\\For more details on scalar quantisation in generic AdS spacetime, see \cite{bianchi2021quantumscalartheoriesadscft}. 

\section{The Flat-Spacetime Limit}\label{flat}

We will first check if we can retrieve the flat-space GAdS master equation from \eqref{eq:lindblad_ads}. First, we note that the flat-space graviton can be written in terms of the spherical polar coordinates as in Ref.~\cite{OnigaWang2016},
\begin{align}
h_{ij}(t,r,\theta,\phi)
&=
4\pi
\sum_{\lambda,\ell,m}
\int dk\,k^2
\sqrt{\frac{G\hbar}{\pi^2 c^2 \omega_k}}
\,
(-i)^\ell
j_\ell(kr)
Y_{\ell m}^{*}(\theta,\phi)\nonumber\\
&\times G^{(\lambda)}_{\ell m,ij}(k)
e^{-i\omega_k t}
+\mathrm{H.c.}
\label{eq:spherical_graviton}
\end{align}
where 
\begin{equation}
G^{(\lambda)}_{\ell m,ij}(k)
=
\int d\Omega_k \,
a_{k,\lambda}
\,e^{(\lambda)}_{ij}(\hat{k})
Y_{\ell m}(\hat{k})\,.
\label{eq:Glm_definition}
\end{equation}
This can be easily derived by substituting the partial wave expansion of the plane waves, $$e^{i\mathbf{k}\cdot\mathbf{x}} =
4\pi\sum_{\ell,m}i^\ell j_\ell(kr) Y_{\ell m}(\hat{x}) Y_{\ell m}^{*}(\hat{k}),$$ where $\hat{x} = (\hat{\theta}, \hat{\phi})$. 

Now, for taking the flat-space limit of the $GAdS_4$ equation, we work in the \textit{confluent energy limit}, that is, we take $n\longrightarrow\infty, \ell\longrightarrow\text{finite}, L\longrightarrow\infty$~\footnote{In principle, the physics of taking the flat-space limit remains the same if $\ell\longrightarrow\infty$, and it can be straightforwardly seen that the result obtained is the same for a finite fixed $\ell$ and the $\ell$ becoming infinity limit.}, but we keep, $$\omega_{n\ell} = c/L(2n + \ell + 2)\longrightarrow \omega_k (\equiv k)= 2nc/L$$
fixed, to keep the energy of the system finite with respect to the flat-space geodesic observer. In this limit, $$\sum_{n\ell}(2\ell + 1)\longrightarrow \int dkd\omega_k\frac{L^3\omega_{k}^2}{2c^3}\equiv \int dkd\vec{n}\frac{L^3\omega_{k}^2}{2c^3},$$ where, we have used Eq.~\eqref{DOS} already. Now we note that,
\begin{equation}
    \int d\vec{n}(P_{11}(\vec{n}))^2 = \frac{32\pi}{15}\,.
    \label{polarization}
\end{equation}
Moreover, in flat space, we will be able to take the Markov approximation, and can thus take the $s$-integral upper limit to infinity in Eq.~\eqref{eq:gads_master_equation_discrete}. Thus, we use another integral identity,
\begin{equation}
\int^\infty_0 ds\text{ }e^{-i(\omega_k - 2\omega_m)s} = \pi\delta(\omega_k - 2\omega_m) \,.
\label{delta}
\end{equation}
Finally, we look at the GAdS radial function $R_{n\ell}(\rho)$ given by \eqref{mq}. By the definition of $\rho$, we have, for large $L$, $r = L\rho$ for finite $r$, which is the radial coordinate in the Schwarzschild chart of the matter-wave system's mean position. Also, $\rho\longrightarrow 0$. We use Stirling's approximation for large $x$, $\frac{\Gamma(x+a)}{\Gamma(x+b)} = x^{a-b}$, $n+\ell \approx n$ for $N_{n\ell}$ to finally find,
\begin{equation}
    N^2_{n\ell} = 2.n^{-3/2}.n^{3/2} = 2\,.
\end{equation}
Now, $\sinh(\rho) \approx \rho\implies (\sinh\rho)^\ell\approx (r/L)^\ell$. Also, $\cosh(\rho)\approx 1\implies (\cosh\rho)^{-2}\approx 1$. 
Now, for the hypergeometric part, with large-$n$ and small-$\rho$ limit, we note,
\begin{align}
{}_2F_1\!\left(
-n,\,
n+\ell+2;\,
\ell+\frac{3}{2};\,
\tanh^2\rho
\right)\nonumber\\
\;\longrightarrow\;
{}_0F_1\!\left(
;\,
\ell+\frac{3}{2};\,
-\frac{k^2 r^2}{4}
\right)\,.
\label{eq:hypergeometric_flat_limit}
\end{align}
We now note the relation between ${}_0F_1$ to spherical Bessel functions,
\begin{equation}
{}_0F_1\!\left(
;\nu+1;
-\frac{x^2}{4}
\right)
=
\Gamma(\nu+1)
\left(
\frac{x}{2}
\right)^{-\nu}
J_\nu(x)\,.
\label{eq:0F1_bessel_identity}
\end{equation}
We note, $\nu = \ell + 1/2$. Thus, we finally find,
\begin{equation}
{}_0F_1\!\left(
;
\ell+\frac{3}{2};
-\frac{k^2 r^2}{4}
\right)
=
\Gamma\!\left(
\ell+\frac{3}{2}
\right)
\left(
\frac{kr}{2}
\right)^{-\ell-\frac12}
J_{\ell+\frac12}(kr)\,.
\label{eq:0F1_spherical_bessel}
\end{equation}
We also note,
\begin{equation}
j_\ell(x)
=
\sqrt{\frac{\pi}{2x}}
\,J_{\ell+\frac12}(x)
\label{eq:spherical_bessel_relation}
\end{equation}
Thus, up to some dimensionless finite constants, we find,
\begin{equation}
R_{n\ell}(\rho)
\sim
\frac{ N_{n\ell}
}{
\pi L^{3/2}
}
\left(
\frac{r}{L}
\right)^\ell
\frac{
2^{\ell+1} \Gamma\!\left(\ell+\frac32\right)
}{
\sqrt{\pi}
}
(kr)^{-\ell}
j_\ell(kr)\,.
\label{eq:ads_radial_flat_limit}
\end{equation}
Therefore, the $r^\ell$ cancels out, keeping a $L^{-\ell}$ in the coefficient of the limiting hypergeometric function. Apparently, the term becomes divergent. 
However, since there is no conserved 4-momentum direction in GAdS spacetime - because the system is like a finite box - we have to redefine the creation and annihilation graviton operators in the flat-space limit, relating to the flat-space creation and annihilation operators written in Cartesian coordinates to those of the GAdS spherical polar coordinates. That relation can easily be written down by looking at \eqref{eq:ads_radial_flat_limit} and \eqref{eq:spherical_graviton} as,
\begin{equation}
a_{n\ell m\lambda}\,
\epsilon^{(\ell,m,\lambda)}_{ij}
\sim
4k^{2}\frac{(\pi L)^{3/2}}{2}
\,N_{\ell}^{-1}
(-i)^{\ell}
Y^{*}_{\ell m}
\,G^{(\lambda)}_{\ell m,ij}(k)\,,
\label{eq:ads_mode_relation}
\end{equation}
with $N_\ell = \frac{2^{\ell + 1}\Gamma(\ell + 3/2)}{\sqrt{\pi}}(kL)^{-\ell}$. Hence, the field operator normalisation changes the divergent term in the graviton field renormalisation in the Cartesian coordinates. Hence, reverting back to Cartesian coordinates, we find Eq.\eqref{eq:gads_master_equation_discrete} turns out to be 
\begin{widetext}
    \begin{equation}
\frac{d\rho_t}{dt}
=
-
\int_0^{\infty} d\omega_{k}\,
\frac{
G\hbar\omega_{k}^{5}
}{
30\pi c^{5}\omega_m^{2}
}
\int_0^{\infty} ds
\Big[
e^{-i(\omega_{k}-2\omega_m)s}
\left(
b^{2}b^{2}\rho_t
-
b^{2}\rho_t b^{2}
\right)
+
e^{i(\omega_{k}-2\omega_m)s}
\left(
\rho_t b^{2}b^{2}
-
b^{2}\rho_t b^{2}
\right)
\Big]\,
\label{eq:gravitational_master_equation_frequency}
\end{equation}
\end{widetext}
where we have already used Eq.~\eqref{polarization}. Keeping only the positive frequency terms (as real gravitons cannot have negative frequency) and performing the delta-integral \eqref{delta}, we finally find the simple Lindblad equation,
\begin{equation}
    \frac{d}{dt}\rho_t = \gamma_{grav}(b^2\rho_tb^{\dagger2} - \frac{1}{2}\{b^{\dagger2}b^2,\rho_t \}); \gamma_{grav} = \frac{32}{15}t^2_{Pl}\omega_m^3\,,
\end{equation}
which exactly corresponds to equation Eq.(48) of \cite{Toro__2024}. 
Note that the apparent divergence in Eq.~\eqref{eq:ads_mode_relation} as $L\longrightarrow\infty$ is not something to get worried about, as the method of canonical quantisation is not generally covariant, and the mode coefficients' expansion is defined such that it compensates exactly the divergence that is created in Eq.~\eqref{eq:ads_radial_flat_limit} in the flat-space limit. In general, the stress-energy tensor associated with the gravitational field confined in a finite volume (which may be used as an alternate quantification of the state of the local description of the quantum gravitational field, see \cite{Fareghbal_2019}) diverges due to the boundary effects, which is reflected in the apparent divergence we found as well. A way to resolve this problem is discussed briefly in Appendix \ref{HoCo}. In our analysis, we did not consider holographic boundary effects on the Von Neumann equation, as we were only dealing with a zeroth-order approximation, but it would be an interesting exercise to see whether including these boundary counterterms in the total Hamiltonian yields a smoother matching to the flat-space equation. However, we will not address this aspect here and will postpone it for a future discussion.

\section{Analogous Lamb Shift for AdS spacetime}\label{Lamb}

The second term of Eq.~\eqref{eq:lindblad_ads}, that is, the term multiplied by $i\alpha_I$ is actually a shift in the vacuum energy GAdS. To see this, we first note that $b^{\dagger2}b^2$ is a Hermitian operator, so it can be expressed as an effective Hamiltonian $H_{eff}$. Then, the term effectively becomes $[\rho, H_{eff}]$, which is analogous to the Lamb shift operator in cavity QED (for more details, see Ref.~\ref{CQED}). We interpret this as a virtual graviton coupling to the matter-wave system, mediated by the graviton field (arising from vacuum fluctuations of the quantum gravitational field), which effectively induces an energy shift relative to the observable field. The number operator for the matter-wave system is given by $\hat{n} = b^\dagger b$. Compared with cavity QED, we have two nonlinear quantum Kerr shifts, yielding an intensity-dependent energy correction. 

Like the Lamb-Shift operator in cavity QED, this term is actually bounded, oscillatory, and reversible - unlike a Casimir energy, thus in principle observable \cite{Grundler2013}. 

\section{A Lightening Brief Note on Cavity QED}\label{CQED}
We define the total system comprising of a 2-level Jaynes-Cummings model and two independent harmonic oscillator baths - one representing the external electromagnetic modes (cavity leakage) and the other representing the background vacuum modes (atomic spontaneous emission). The total Hamiltonian is given by, 
\begin{equation}
H = H_{JC} + H_{bath} + H_{int}\,.
\end{equation}
The Jaynes-Cummings Hamiltonian is given by ($\hbar = 1$ and $\sigma_x, \sigma_y, \sigma_z$ are Pauli matrices with $\sigma_+, \sigma_-$ being the corresponding raising and lowering operators created out of them) see \cite{BerciuJC},
\begin{equation}
    H_{JC} = \omega_c a^\dagger a + \frac{\omega_a}{2}\sigma_z + g(\sigma_+a + \sigma_-a^\dagger)\,. 
\end{equation}
The bath Hamiltonian is given by ($b_k$ are operators for cavity loss environment, whereas $c_j$ are for atomic decay environment),
\begin{equation}
    H_{bath} = \sum_k\omega_kb_k^\dagger b_k + \sum_j \nu_jc_j^\dagger c_j\,.
\end{equation}
The interaction Hamiltonian will be given as,
\begin{equation}
    H_{int} = \sum_{k}(g_ka^\dagger b_k + g_k^*ab_k^\dagger) + \sum_j (\lambda_j\sigma_+c_j + \lambda^*_j\sigma_- c_j^\dagger)\,.
\end{equation}
It is clearly visible that in this particular form of the interaction Hamiltonian, it is difficult to calculate the non-Markovian memory kernel - if we consider the cavity to be finite-sized, then the Rabi oscillations cannot be considered the whole picture in this scenario where we are treating the cavity being coupled to an external bath of modes to model cavity leakage - in principle, this is closer to the "real cavity" scenario. The Markov approximation will not be valid in the derivation of the master equation. Instead, if we model the cavity as a multi-mode system, and modify the Jaynes-Cummings Hamiltonian to include the interaction (as in Wigner-Weisskopf theory, where we build the cavity physics by treating the external field along with it simultaneously, such that the bath becomes a non-trivial reservoir), we will have a Hamiltonian given by,
\begin{equation}
    H = \omega_a \sigma_+ \sigma_- + \sum_k \omega_k b_k^\dagger b_k + \sum_k (g_k \sigma_+ b_k + g_k^* \sigma_- b_k^\dagger)\,.
\end{equation}
The cavity structure becomes encoded in the spectral density of the reservoir with leakage rate $\kappa$ and $W$ being the effective overall coupling strength to the cavity mode, which is a Lorentzian distribution,
\begin{equation}
    J(\omega) = \frac{1}{2\pi} \frac{W^2 \kappa}{(\omega - \omega_c)^2 + (\kappa/2)^2}\,.
    \label{Lorentzian}
\end{equation}
Assuming the bath is initially in the vacuum state $\ket{0}_B$, and the atom is in a superposition, the total state resides in a single-excitation subspace. We can write the exact state vector as
\begin{equation}
    |\psi(t)\rangle = c_e(t) e^{-i\omega_a t} |e, 0\rangle_B + \sum_k c_k(t) e^{-i\omega_k t} |g, 1_k\rangle_B\,.
\end{equation}
Plugging into the Schrödinger equation in the interaction picture, we get a system of coupled differential equations for the amplitude, 
\begin{align}
    \dot{c}_e(t) = -i \sum_k g_k e^{i(\omega_a - \omega_k)t} c_k(t)\nonumber\\
    \dot{c}_k(t) = -i g_k^* e^{-i(\omega_a - \omega_k)t} c_e(t)\,.
\end{align}
We formally integrate $\dot{c}_k(t)$ from 0 to $t$ (assuming $c_k(0) = 0$) and substitute it back in equation for $\dot{c}_e(t)$ to get,
\begin{equation}
    \dot{c}_e(t) = -\int_0^t dt' \sum_k |g_k|^2 e^{i(\omega_a - \omega_k)(t - t')} c_e(t')\,.
\end{equation}
We can now replace the sum over modes by an integral over the spectral density modes, thus yielding,
\begin{equation}
    \dot{c}_e(t) = -\int_0^t dt' f(t - t') c_e(t')
\end{equation}
Here, $f(t-t')$ is the memory kernel, which is given by
\begin{equation}
    f(t - t') = \int d\omega J(\omega) e^{i(\omega_a - \omega)(t - t')}\,.
    \label{memory}
\end{equation}
To find a master equation of the reduced density matrix of the atom, $\rho_{JC}(t) = \text{Tr}_B \{ |\psi(t)\rangle \langle\psi(t)| \}$, we can find the exact solution of the amplitude in the form,
\begin{equation}
    \dot{c}_e(t) = \left[ -\frac{\gamma(t)}{2} - i S(t) \right] c_e(t)\,.
\end{equation}
Because $\rho_{JC}(t)$ in the basis $(\ket{e}, \ket{g})$ is fully determined by $c_e(t)$ (such that $\rho_{ee}(t) = |c_e(t)|^2$ and $\rho_{eg}(t) = c_e(t)c_g^*(0)$), we can take the time derivative of $\rho_S(t)$ and directly substitute to find the exact non-Markovian master equation in the Schrödinger picture,
\begin{align}
    \dot{\rho}_{JC}(t) = -i \frac{S(t)}{2} [\sigma_+ \sigma_-, \rho_{JC}(t)] + \gamma(t) ( \sigma_- \rho_{JC}(t) \sigma_+\nonumber\\
    - \frac{1}{2} \{ \sigma_+ \sigma_-, \rho_{JC}(t) \})\,.
    \label{cavity_qed_master}
\end{align}
The time-dependent decay rate (replacing the Lindbald dissipator) $\gamma(t)$ and the Lamb shift $S(t)$ are defined by,
\begin{equation}
    \gamma(t) = -2 \text{Re} \left[ \frac{\dot{c}_e(t)}{c_e(t)} \right], \quad S(t) = -\text{Im} \left[ \frac{\dot{c}_e(t)}{c_e(t)} \right]\,.
\end{equation}
For the Lorentizan spectral density in \eqref{Lorentzian} and assuming resonance, $\omega_a = \omega_c$, we can easily see that the memory kernel defined in \eqref{memory} turn out to be ($\tau = t-t'$), 
\begin{equation}
    f(\tau) = \frac{W^2}{2} e^{-\kappa \tau / 2}
\end{equation}
Solving the integro-differential equation of the amplitude with this kernel yields the solution,
\begin{equation}
    c_e(t) = e^{-\kappa t / 4} \left[ \cosh\left(\frac{dt}{4}\right) + \frac{\kappa}{d} \sinh\left(\frac{dt}{4}\right) \right]\,,
\end{equation}
with $d = \sqrt{\kappa^2 - 8W^2}$. From this exact amplitude, the time-dependent decay rate can be found out to be,
\begin{equation}
    \gamma(t) = \frac{4W^2 \sinh(dt/4)}{d \cosh(dt/4) + \kappa \sinh(dt/4)}\,.
    \label{decay}
\end{equation}
If $8W^2 < \kappa^2$, $d$ is always real and therefore, $\gamma(t)$ approaches smoothly an asymptotic value (the Purcell effect) called the Markov limit, $\gamma_M = 4W^2/\kappa$. However, if $8W^2 > \kappa^2$ - then $d$ becomes imaginary, prompting the hyperbolic functions in \eqref{decay} to turn into trigonometric functions, and thus, $\rho_{JC}(t)$ oscillates as per \eqref{cavity_qed_master}. Specifically, $\gamma(t)$ becomes negative for certain time intervals - and a negative decay rate in the dissipator indicates a temporary reversal of the environmental action. Thus, recoherence occurs in alternation to decoherence, where quantum information is flowing back from the cavity into the atom - this is the signature of non-Markovian systems. $S(t)$ accounts for the time-dependent Lamb shift. Because the system is strongly coupled to a structured bath, the effective energy level of the atom is shielded by the EM field in the cavity, leading to a dynamics transformation via its transition frequency. It is clear that for this particular Lorentzian distribution, $S(t) = 0$ if $d$ is real. If $d = i\Omega$ is completely imaginary, then the Lamb-shift term will be given by,
\begin{equation}
    S(t) = \frac{\kappa\Omega^2}{4[\Omega^2\cos^2(\frac{\Omega t}{4}) + \kappa^2\sin^2(\frac{\Omega t}{4})]}\,.
\end{equation}
More details can be found in \cite{Miao_2025}. 

\section{\texorpdfstring{A Holographic Counterterm in the $AdS$-action}{A Holographic Counterterm in the AdS-action}}\label{HoCo}
The quasilocal Brown-York stress-energy tensor associated with a generally covariant spacetime locally on the boundary is given by
\begin{equation}
    T_{\mu\nu} = \frac{2}{\sqrt{-\gamma}}\frac{\delta S_{grav}}{\delta \gamma^{\mu\nu}}\,.
\end{equation}
Here, $\gamma_{\mu\nu}$ is the metric on the boundary. The resulting stress-energy tensor diverges as the boundary of the spacetime (in our case, AdS) is taken to infinity. In AdS, the way to bypass this is to add local counterterms to the boundary action of gravity, because in light of the AdS/CFT correspondence, these divergences are just the standard ultraviolet infinities that arise in the quantum CFT residing on the boundary of the spacetime. These counterterms are free of the ambiguities that arise when embedding the boundary in a reference spacetime. 
\\From an ADM-foliation decomposition of a $(d+1)$-dimensional AdS spacetime $\mathcal{M}$ by a series of $d$-dimensional timelike surfaces homeomorphic to the boundary $\partial\mathcal{M}$ spanned by the coordinates $x^\mu$ and $r$ being the remaining coordinate, we can write the metric as,
\begin{equation}
    ds^2 = N^2dr^2 + \gamma_{\mu\nu}(dx^\mu + N^\mu dr)(dx^\nu + N^\nu dr)\,,
\end{equation}
where $\gamma_{\mu\nu}$ is a function of all coordinates, and it is the metric evaluated at a fixed $r$ on $\partial\mathcal{M}_r$\footnote{The AdS boundary is generally considered as part of a conformal class.}. Varying the gravitational action will only be contributed by the boundary terms (as the EOM kills the bulk terms) and will imply the quasilocal stress-energy tensor as, 
\begin{equation}
    T^{\mu\nu} = \frac{1}{8\pi}[\Theta^{\mu\nu} - \Theta\gamma^{\mu\nu} + \frac{2}{\sqrt{-\gamma}}\frac{\delta S_{ct}}{\delta\gamma_{\mu\nu}}]\,.
\end{equation}
We note that,
\begin{equation}
    \Theta_{\mu\nu} = -\frac{1}{2}(\nabla_\mu n_{\nu} + \nabla_\nu n_\mu)\,,
    \label{extrinsic}
\end{equation}
where, $n^\mu$ is the normal vector to boundary $\partial\mathcal{M}_r$. The counterterm action $S_{ct}$ is chosen such that it cancels the divergences that arise as $\partial\mathcal{M}_r$ approaches the AdS horizon $\partial\mathcal{M}$. 

To calculate $S_{ct}$, the approach is to compute the ADM mass~\cite{} associated with the spacetime metric and require its finiteness. To define the former, we take a spacelike foliation of the AdS metric, multiply the lapse function by the proper energy density computed using the timelike unit normal to the hypersurface and the stress-energy tensor defined on the same hypersurface, and then integrate over the whole hypersurface. It can be tediously but straightforwardly shown that the counterterms coming from combining the contributions from both the Poincaré patch of the AdS and the global chart of it for $AdS_4$ are given by (with $S_{ct} = \int_{\partial\mathcal{M}_r}L_{ct}$),
\begin{equation}
    L_{ct} = -\frac{2}{L}\sqrt{-\gamma}(1 - \frac{L^2}{4}R)\,,
\end{equation}
where, $R$ is the Ricci scalar of $\gamma_{\mu\nu}$. Note that the mass contribution differs between the Poincaré and global charts of the AdS spacetime - but that is only to be expected, as this mass is observer-dependent, and therefore, since the time direction in these two metrics differs, they have different definitions of the ADM mass. 
\\All the things written in this appendix - and more details - can be found in \cite{Balasubramanian_1999}. 

\section{The Matter-Gravitational Field Interaction Term at the Vertex Level}\label{APPD}

In general, a general relativistic point-particle Lagrangian is given by,
\begin{equation}
    L = -mc^2\sqrt{-g_{\mu\nu}(x)\,\dot x^\mu \dot x^\nu}\,,
\end{equation}
where, $g_{\mu\nu} = \eta_{\mu\nu} + h_{\mu\nu},
\qquad |h_{\mu\nu}| \ll 1$. Now, $-g_{\mu\nu}\dot x^\mu \dot x^\nu
= -\eta_{\mu\nu}\dot x^\mu \dot x^\nu - h_{\mu\nu}\dot x^\mu \dot x^\nu$. If we consider non-relativistic speed, then we have, $\dot x^\mu = (c,\mathbf v), \qquad |\mathbf v| \ll c\implies -\eta_{\mu\nu}\dot x^\mu \dot x^\nu = c^2 - v^2\approx c^2$. So, eventually, we get, $\sqrt{c^2 - h_{\mu\nu}\dot x^\mu \dot x^\nu}\approx c\left[1 - \frac{1}{2c^2} h_{\mu\nu}\dot x^\mu \dot x^\nu\right]$. Thus, effectively, we get, $L
= -mc^2 + \frac{m}{2}h_{\mu\nu}\dot x^\mu \dot x^\nu +O(h^2, v^2/c^2)$. The first term is an irrelevant constant.
\\For a point-particle, the EM-tensor is given by,
\begin{equation}
    T^{\mu\nu}(x) = m \int d\tau\; \dot x^\mu \dot x^\nu\, \delta^{(4)}\!\big(x - x(\tau)\big)\,.
\end{equation}
Thus, the interaction becomes, $S_{\text{int}} = -\frac{1}{2}\int d^4x\;h_{\mu\nu}(x)\,T^{\mu\nu}(x)$. Effectively, the interaction Hamiltonian becomes,
\begin{equation}
    H_{\text{int}} = \frac{1}{2} \int d^3x\;h_{\mu\nu}(\mathbf x,t)\, T^{\mu\nu}(\mathbf x,t)\,.
\end{equation}

\bibliographystyle{apsrev4-2}
\bibliography{refs}

\end{document}